\documentclass[reprint,amsmath,amssymb,aps,prb]{revtex4-1}

\pdfoutput=1

\usepackage{hyperref,url}
\usepackage{amsmath}
\usepackage{amsfonts}
\usepackage[capitalise]{cleveref}
\usepackage{placeins}
\hypersetup{breaklinks,colorlinks=true,linkcolor=blue,citecolor=blue,urlcolor=blue,filecolor=blue}
\usepackage{graphicx}
\usepackage{dcolumn}
\usepackage{algorithm}
\usepackage{algpseudocode}

\newcommand\mat\mathbf
\newcommand\tr{\operatorname{tr}}

\newcommand{\joonho}[1]{{}}
\newcommand{\insertnew}[1]{{\textcolor{black}{#1}}}
\newcommand{\insertrev}[1]{{\textcolor{black}{#1}}}

\begin{document}

\author{Joonho Lee}
\email{linusjoonho@gmail.com}
\affiliation{Department of Chemistry, Columbia University, New York, NY 10027, USA.}
\author{Miguel A. Morales}
\affiliation{Quantum Simulations Group, Lawrence Livermore National Laboratory, 7000 East Avenue, Livermore, CA, 94551 USA.}
\author{Fionn D. Malone}
\email{malone14@llnl.gov}
\affiliation{Quantum Simulations Group, Lawrence Livermore National Laboratory, 7000 East Avenue, Livermore, CA, 94551 USA.}

\title{A Phaseless Auxiliary-Field Quantum Monte Carlo Perspective on the Uniform Electron Gas at Finite Temperatures:
Issues, Observations, and Benchmark Study
}
\begin{abstract}
We investigate the viability of the phaseless finite temperature auxiliary field quantum Monte Carlo (ph-FT-AFQMC) method for ab initio systems using the uniform electron gas as a model.
Through comparisons with exact results and finite temperature coupled cluster theory, we find that ph-FT-AFQMC is sufficiently accurate at high to intermediate electronic densities.
We show both analytically and numerically that the phaseless constraint at finite temperature is fundamentally different from its zero temperature counterpart (i.e., ph-ZT-AFQMC) and generally one should not expect ph-FT-AFQMC to agree with ph-ZT-AFQMC in the low temperature limit.
With an efficient implementation, we are able to compare exchange-correlation energies to existing results in the thermodynamic limit and find that existing parameterizations are highly accurate. \insertnew{In particular, we found that ph-FT-AFQMC exchange-correlation energies are in a better agreement with a known parametrization than is restricted path-integral Monte Carlo in the regime of $\Theta\le0.5$ and $r_s \le 2$, which highlights the strength of ph-FT-AFQMC.}
\end{abstract}
\maketitle
\section{Introduction}
Temperature-dependent properties of interacting fermions are fundamentally important in both experiments and theory.
Typical phenomena at finite temperature include the BCS-BEC crossover in the attractive 2D Fermi model,\citep{PetrovBEC2003} the competition between stripe and superconducting orders in the two-dimensional Hubbard model,\citep{QinSuperconductivity2020} and plasmonic catalysis.\citep{MukerjeePlasmonic2013}
Furthermore, there is a growing interest in warm-dense matter,\citep{GrazianiFrontiers2014} an extreme state of matter found in planetary interiors\citep{FortneyPlanets2009} that can be created with high intensity lasers.\citep{KoenigFusion2005,ErnstorferGold2009}
Understanding such phenomena at the theoretical level is challenging due to the delicate interplay between electron-electron, electron-ion, quantum mechanical and thermal effects all of which can be equally important and often cannot be treated perturbatively.
Thus, accurate computational approaches are required that are capable of capturing these effects.

Density functional theory (DFT), as the workhorse of zero temperature electronic structure theory, is an ideal candidate: it is relatively cheap and accurate, it can be coupled with molecular dynamics to include ionic effects and it can be rigorously formulated to incorporate thermal electronic effects.\citep{MerminTDFT1965}
Indeed, DFT has proven itself effective in simulating warm dense matter.\citep{DriverOxygen2015,ZhangSodium12016,ZhangSodium22017,DriverRev2017}
However, questions remain regarding the accuracy of using approximations made in thermal DFT, including the accuracy of the use of zero temperature exchange correlation functionals.\citep{KarasievImportance2016,KarasievDeuterium2019,DornheimXC2020,MihaylovHybrid2020}
Complementary approaches are therefore desired that can benchmark or supplement DFT results when necessary.

Quantum Monte Carlo (QMC) methods are a promising class of such computational methods.
They are in principle exact in a finite supercell and can explicitly include many-body and thermal effects in an unbiased way.
Of the many flavors of QMC, real space path integral Monte Carlo (PIMC) is perhaps the best known and well established.\citep{ceperley_pimc}
PIMC, as a real space approach, works in the complete basis set limit which is a considerable benefit at very high electronic temperatures. However, like all fermionic QMC methods it suffers from the sign problem which can only be overcome using the uncontrolled restricted path approximation,\citep{ceperley_fermion_nodes} which leads to the restricted PIMC (RPIMC) approach.
This approach is similar in spirit to the fixed-node approximation in diffusion Monte Carlo (DMC),\citep{foulkes_quantum_2001} but now a constraint is enforced using a trial thermal density matrix. The quality of this constraint is a priori unknown, however, results with free fermion nodes in the uniform electron gas (UEG) suggest that it is unreliable at high densities and at lower temperatures.\citep{brown_path-integral_2013,schoof_textitab_2015,malone_accurate_2016}
We note there have been promising developments in extending the scope of PIMC to higher densities and lower temperatures through algorithmic developments\citep{dubois_sign} including the development of new algorithms such as  \insertnew{the in principle exact} real space permutation blocking PIMC\citep{dornheim_pbpimc} (PB-PIMC) and second quantized configuration PIMC\citep{schoof_cpimc} (CPIMC). \insertnew{In particular, PB-PIMC and CPIMC have been used as complementary approaches to simulate the warm dense UEG above half the Fermi temperature, with the fermion sign problem preventing simulations below this.}
\insertrev{We note that restricted CPIMC may be a promising route to access lower temperature given its results on small UEG supercells and the 33-electron supercell spin-polarized case.\cite{Yilmaz2020Sep}}
Other interesting QMC approaches  
are the quantum chemistry inspired methods like density matrix quantum Monte Carlo\citep{blunt_density-matrix_2014,malone_interaction_2015,PetrasDMQMC2020} (DMQMC) and krylov-projected full configuration interaction quantum Monte Carlo.\citep{blunt_krylov-projected_2015}
Like CPIMC these approaches work in a second quantized space and work well at high densities, however they often struggle to reach the complete basis set limit.
All of these methods offer unbiased exact thermal expectation values in principle, however, they all ultimately scale exponentially with the number of electrons in general.
\insertnew{See Ref.\citenum{dornheim_qmc_review} for a review of the parameter regimes accessible to these new QMC methods.}

Recently there has also been considerable interest in developing finite temperature deterministic quantum chemistry methods
that work in a finite basis set. These methods include
second order perturbation theory,\citep{He2014,Santra2016,Hirata2020,Jha2020} coupled cluster theory\citep{Hermes2015,White2018,White2020} and thermofield theory.\citep{harsha_thermofield_2019}
These are promising and offer a systematic approach to including electronic temperature effects, 
but they often struggle to reach the continuum limit (also known as the complete basis set (CBS) limit)
due to the steep computational scaling with respect to the number of basis functions, $M$.
%
For example, FT-CCSD scales like $\mathcal O(M^6)$, which becomes prohibitively expensive to converge results to the CBS limit.\citep{White2018,White2020}

Finite temperature auxiliary-Field QMC\citep{blankenbecler_dqmc_1,scalapino_dqmc} (FT-AFQMC) is another promising QMC method. 
It works in the second quantized space and thereby suffers from the basis set incompleteness errors common to DMQMC \insertnew{and CPIMC}. 
However, unlike DMQMC \insertnew{and CPIMC}, it can be made to scale polynomially with system size at the cost of introducing an uncontrolled bias called the phaseless approximation.\citep{zhang_phaseless,ZhangFTAFQMC1999}
Moreover, zero temperature phaseless AFQMC (ph-ZT-AFQMC) has proven itself as one of the most accurate and scalable post Hartree--Fock methods.\citep{simons_hubbard_2d,motta_hydrogen,williams_transition_metal_simons,lee2020performance,lee2020utilizing}
In addition, the bias introduced by the phaseless approximation is typically much smaller than the fixed node-error in DMC\citep{Malone2020} which in principle should serve as an rough upper bound to the bias in RPIMC.
Thus, it is important to assess the quality of ph-FT-AFQMC for realistic systems as to date the applications have largely focussed on model systems\citep{HeSCFT2019,RubensteinBose2012} or have not enforced the constraint\citep{LiuFTAI2018,Liu2020,Shen2020} which is not a practical approach as the system size increases.
Compared to FT-CCSD, ph-FT-AFQMC maintains favorable cubic scaling $\mathcal O(M^3)$ for each statistical sample,\cite{lee2020stochastic,malone_isdf} which makes it better-suited for large-scale warm dense matter simulations.

In this paper, we investigate the viability of ph-FT-AFQMC both in terms of its accuracy and in its ability to reach the complete basis set limit.
We stress good performance for \emph{both} of these metrics is critical if the method is to be practically useful for finite-temperature ab-initio systems.
To address these problems we use the uniform electron gas (UEG) model as a testbed for ph-FT-AFQMC.
To the best of our knowledge, 
our work is the first to apply 
ph-FT-AFQMC at finite temperature beyond lattice models with short-range interactions. 
The main goal of our work is to assess
the accuracy of ph-FT-AFQMC when applied to the UEG model.

Apart from being a foundational model in condensed matter physics,\citep{vignale_qtel,ceperley_alder,PerdewZungerUEG1981,PBE1996} the UEG offers a number of useful features from a computational point of view. First, the model can be tuned from weak to strong correlation as a function of the density parameter, $r_s$.
Given that our previous work\citep{LeeUEG2019} has shown that ph-ZT-AFQMC is highly accurate for $r_s\le 3-4$ at zero temperature and we have an idea of the magnitude of errors we might expect.
Second, basis set convergence can be easily investigated in a planewave basis set by increasing the energy cutoff.
Finally, the UEG at warm dense matter conditions has been the subject of intense study over the past decade.\citep{dornheim_review}
In fact many of the recent developments\citep{dornheim_review} in finite temperature fermionic QMC methods was spurred on by a discrepancy between RPIMC and CPIMC\citep{BrownUEG1,schoof_prl} results for the warm dense UEG that have been incorporated into finite temperature exchange correlation functionals.\citep{brown_exchange-correlation_2013-1,KarasievFit2014,Groth2017}
Because of this effort there is a considerable amount of essentially exact data for energetic,\citep{Groth2016,DornheimUnpolarized2016} static\citep{Dornheim2017,DornheimStaticStructure2017,GrothStaticStructure2017} and dynamic properties\citep{Dornheim2018,Groth2019} of the model at a wide range of densities and temperatures.
Nonetheless, despite this immense effort, there is a gap in accurate data below roughly half the Fermi temperature below which no method could reach due to the Fermionic sign problem.\citep{Dornheim2016TDL}
In this paper we use ph-FT-AFQMC to partially fill this gap.

This paper is organized as follows. In the first section we outline the ph-FT-AFQMC method, paying careful attention to how to implement it efficiently in a numerically stable fashion. Next, we carefully benchmark the method at various temperatures against exact results in small basis sets and
investigate the low temperature limit in comparison to the ph-ZT-AFQMC results.
We also compare the ph-FT-AFQMC results to FT-CCSD and PIMC where possible. 
Lastly, we compare our results against other QMC methods
and available excahnge-correlation parametrization in the thermodynamic limit
and finish by outlining our perspective for the future of the method.

\section{Theory}
We briefly summarize the phaseless approximation of AFQMC within the finite temperature formalism (i.e., ph-FT-AFQMC).

\subsection{Finite-Temperature AFQMC in the Grand Canonical Ensemble}
\subsubsection{General formalisms}
The finite temperature AFQMC algorithm aims to compute thermal expectation values based on the grand canonical partition function
\begin{equation}
Z = \tr (e^{-\beta(\hat{H}-\mu\hat{N})})
\label{eq:Z}
\end{equation}
where $\hat{N}$ is a total number operator, $\mu$ is a chemical potential,
and the Hamiltonian involves one-body ($\hat{H}_1$) and two-body ($\hat{H}_2$) terms,
\begin{equation}
    \hat{H} = \hat{H}_1 + \hat{H}_2
\end{equation}
The direct (deterministic) evaluation of trace in \cref{eq:Z}
scales exponentially since there are exponentially many states to consider.
It is then natural to consider QMC algorithms to sample an instance of the terms in $Z$.

A particular flavor of QMC that we focus on here is AFQMC where
the two-body propagator is 
expressed by the Hubbard-Stratonovich transformation,\citep{hubbard_strat}
\begin{equation}
e^{-\Delta \tau\hat{H}_2} = 
\int d\mathbf{x}
p(\mathbf{x})
e^{-\sqrt{\Delta\tau} \mathbf{x}\cdot\hat{\mathbf{v}}}
\label{eq:B}
\end{equation}
where $\hat{\mathbf v}$ is related to
the two-body Hamiltonian as a sum of squared operators,
\begin{equation}
    \hat{H}_2 = -\frac12 \sum_\alpha^{N_\alpha} \hat{v}_\alpha^2,
\end{equation}
$\mathbf x$ is a vector of $N_\alpha$ auxiliary fields
which are samples from the standard normal distribution, $p(\mathbf x)$.
With the symmetric Trotter decomposition, the total propagator reads
\begin{equation}
\exp(-\Delta\tau \hat{H}) \: = 
\int d\mathbf{x}
p(\mathbf{x})
\hat{B}(\Delta \tau, \mathbf x, \mu)
\label{eq:HS}
\end{equation}
where $\hat{B}$ is defined as
\begin{equation}
\hat{B}(\Delta \tau, \mathbf x, \mu) = 
e^{-\frac{\Delta\tau}{2} (\hat{H}_1 - \mu \hat{N})}
e^{-\sqrt{\Delta\tau} \mathbf{x}\cdot\hat{\mathbf{v}}}
e^{-\frac{\Delta\tau}{2} (\hat{H}_1 - \mu \hat{N})}.
\label{eq:B}
\end{equation}

For a given number of imaginary time slices $n$,
we sample the auxiliary fields via MC,
\begin{align}\nonumber
Z =
\sum_{\{\mathbf{x}_1, \mathbf{x}_2, \cdot\cdot\cdot \mathbf{x}_n\}}
&p(\mathbf{x}_1, \mathbf{x}_2, \cdot\cdot\cdot \mathbf{x}_n) \\
&\times
\tr\left({\prod_{i=1}^n\hat{B}(\Delta \tau, \mathbf x_i, \mu)}\right)
\label{eq:grand}
\end{align}
where $p(\mathbf{x}_1, \mathbf{x}_2, \cdot\cdot\cdot \mathbf{x}_n)$ is the probability of sampling a specific path designated by auxiliary fields,
$\mathbf{x}_1, \mathbf{x}_2, \cdot\cdot\cdot \mathbf{x}_n$, and the time step,
$\Delta \tau = \beta/n$, is determined by the temperature and the number of imaginary time slices.
The evaluation of the trace in \cref{eq:grand} is still difficult because it needs to
consider all possible states in the Hilbert space despite the fact that every operator inside the trace is a one-body operator.
One can show analytically that the trace can be written in terms of a determinant
in the grand canonical ensemble,\citep{blankenbecler_dqmc_1,Hirsch1985,DosSantos2003}
\begin{align} \nonumber
Z =
\sum_{\{\mathbf{x}_1, \mathbf{x}_2, \cdot\cdot\cdot \mathbf{x}_n\}}
&p(\mathbf{x}_1, \mathbf{x}_2, \cdot\cdot\cdot \mathbf{x}_n)\\
&\times\det\left({\mathbf{I} + \prod_{i=1}^n\mathbf {B}(\Delta \tau, \mathbf x_i, \mu)}\right)
\label{eq:grand2}
\end{align}
where $\mathbf {B}$ is a matrix representation of $\hat{B}$ in the single-particle basis.
For a later use, we define the product of $\mathbf{B}$ as
\begin{equation}
\mathbf{A} (n\Delta\tau,\{\mathbf x_k\})
= \prod_{i=1}^n\mathbf{B}(\Delta \tau, \mathbf x_i, \mu)
\end{equation}
where
we omit $\mu$ in the argument of $\mathbf A$ for simplicity.
Note that for now we defined $\mathbf A$ only at the end
of the imaginary time propagation.
Later, we will define $\mathbf A$ along the trajectory for an 
arbitrary imaginary time $\tau$.

We are mostly interested in computing expectation values based on the partition function in Eq. \eqref{eq:grand2}, not the partition function itself:
\begin{equation}
\langle\hat{O} \rangle
=\frac{\tr(e^{-\beta(\hat{H}-\mu\hat{N})}\hat{O})}{Z}
\end{equation}
The computation of expectation values can be easily achieved through an importance sampling procedure. 
To see this, we first write
\begin{align} \nonumber
\langle\hat{O} \rangle
& =
\frac{1}Z\sum_{\mathbf X}
\tr(\hat{A}(\mathbf X) \hat{O})\\
& = 
\sum_{\mathbf X}
\frac{\tr(\hat{A}(\mathbf X) \hat{O})}
{\tr(\hat{A}(\mathbf X))} 
\frac{\tr(\hat{A}(\mathbf X))}{Z}
\end{align}
where $\mathbf X$ denotes the set of auxiliary fields along each imaginary path.
The field configurations $\mathbf X$ are sampled by the MC algorithm where
we write expectation values
\begin{equation}
\langle\hat{O} \rangle
=
\frac{
\sum_i w_i O_{L,i}(\mathbf X_i)
}
{
\sum_i w_i
}
\end{equation}
where the local expectation value is defined as
\begin{equation}
O_{L,i}(\mathbf X_i)
=
\frac{\tr(\hat{A}(\mathbf X_i) \hat{O})}
{\tr(\hat{A}(\mathbf X_i))}
\end{equation}
and
walker weights $w_i$ 
are
updated
via the importance sampling procedure
based on the
distribution, ${\tr(\hat{A}(\mathbf X))}/{Z}$.

In AFQMC, any expectation values are expressed as a function of the one-body Green's function
\begin{equation}
G_{ij} = \langle \hat{c}_i \hat{c}_j^\dagger\rangle,
\end{equation}
following the generalized Wick's theorem.\citep{bruus2004many} Therefore, computing $\mathbf G$ is sufficient to compute any expectation values.
In terms of $\mathbf{B}(\Delta\tau,\mathbf x, \mu)$, 
it can be shown that 
a sample of 
the one-body Green's function (i.e., the local quantity) is\citep{Hirsch1985,DosSantos2003}
\begin{equation}
\mathbf G(n\Delta\tau,\{\mathbf x_k\}) = 
\left({\mathbf{I} + 
\mathbf{A} (n\Delta\tau,\{\mathbf x_k\}
}\right))^{-1}.
\label{eq:green}
\end{equation}
We note that a sample of the one-body reduced density matrix (1RDM), $\mathbf P$, is obtained by 
\begin{equation}
    \mathbf P (n\Delta\tau,\{\mathbf x_k\}) 
    = \mathbf I - (\mathbf G(n\Delta\tau,\{\mathbf x_k\}))^\text{T}.
\end{equation}
Using Wick's theorem, higher order reduced density matrices can be computed as products of the one-body Green's function.

\subsubsection{Phaseless approximation}
Since the value of determinants in Eq. \eqref{eq:grand2} or equivalently $\tr(\hat{A}(\mathbf X))$ for a given $\mathbf X$ can be either positive or negative (in our case complex),\citep{Wilson1995} the phase problem 
naturally arises in the finite temperature algorithm. Therefore, it is important to impose the phaseless constraint\citep{ZhangFTAFQMC1999,RubensteinBose2012} to remove the phase problem similar to the case of the ph-ZT-AFQMC algorithm.\citep{zhang_phaseless} This way, we can keep the overall scaling of the algorithm polynomial.

We introduce a trial density matrix $\hat{B}_T(\Delta\tau, \mu_T)$ and this is used to impose a phaseless constraint in the imaginary time evolution,\citep{ZhangFTAFQMC1999}
which is defined as
\begin{equation}
\hat{B}_T(\Delta\tau, \mu_T) \equiv e^{-\Delta\tau(\hat{H}_T -\mu_T\hat{N})}
\end{equation}
where $\hat{H}_T$ is some one-body operator and $\mu_T$ is the chemical potential for the trial density matrix.
Throughout the paper,
we will assume that 
$\mu_T$ is tuned so that
the thermal one-body density matrix from $\hat{B}_T(n\Delta\tau, \mu_T)$ has a trace of $N$
with $N$ being the desired number of particles.

We implement the imaginary time evolution of a path by the following algorithm.
The central quantity in the constrained evolution is $\mathbf A$. At an imaginary time $\tau = k\Delta\tau$, we define 
\begin{align}\nonumber
\mathbf{A}(\tau, \{\mathbf x_i\}_{i=1}^{k})
= &
\left(\mathbf{B}_T(\Delta\tau, \mu_T)\right)^{n-k-1}
\mathbf{B}_T(\Delta\tau, \mu_T)\\
&\prod_{i = 1}^{k}
\left(
\mathbf{B}(\Delta\tau,\mathbf x_i,\mu)
\right)
\end{align}
At each imaginary time step, we replace $\mathbf B_T$ in the middle with $\mathbf{B}$. That is,
\begin{align}\nonumber
\mathbf{A}(\tau+\Delta\tau,\{\mathbf x_i\}_{i=1}^{k+1})
= &
\left(\mathbf{B}_T(\Delta\tau, \mu_T)\right)^{n-k-1}
\mathbf{B}(\Delta\tau,\mathbf x_{k+1},\mu)\\
&\times \prod_{i = 1}^{k}
\left(
\mathbf{B}(\Delta\tau,\mathbf x_i,\mu)
\right)
\label{eq:Abrenda}
\end{align}
In general, the product of $\mathbf B$ and/or $\mathbf B_T$ requires a special care for numerical stability especially when simulating low temperature. This can be achieved by the stabilization method described in Ref.\citenum{tomas2012advancing}, which we will describe further in \cref{sec:stab}.

During this propagation, the importance function is determined by 
the FT-AFQMC overlap ratio
\begin{equation}
S_{k}(\tau,\Delta\tau,\{\mathbf x_i\}_{i=1}^{k}) 
= \frac
{\det(\mathbf I + \mathbf{A}(\tau+\Delta\tau,\{\mathbf x_i\}_{i=1}^{k}))}
{\det(\mathbf I + \mathbf{A}(\tau,\{\mathbf x_i\}_{i=1}^{k-1}))}
\label{eq:ftoratio}
\end{equation}
For a later use, 
we
mention that we can equivalently write this ratio in terms of trace over all possible states in the grand canonical ensemble:
\begin{equation}
S_{k}(\tau,\Delta\tau,\{\mathbf x_i\}_{i=1}^{k}) 
= \frac
{\tr(\hat{A}(\tau+\Delta\tau,\{\mathbf x_i\}_{i=1}^{k}))}
{\tr(\hat{A}(\tau,\{\mathbf x_i\}_{i=1}^{k-1}))}
\label{eq:ftoratio2}
\end{equation}
In practice, we employ the ``optimal'' force bias which is a shift to the Gaussian distribution as well
as the mean-field subtraction which enforces the normal ordering.
The optimal force bias at $\tau = k \Delta\tau$ is
\begin{equation}
    \mathbf{\bar{x}}_k = -\sqrt{\Delta\tau}
\sum_\alpha
\left(\sum_{pq}
P_{pq}(\tau)
v_{pq}^{\alpha} - \bar{v}_0^{\alpha}
\right)
\end{equation}
and the mean-field subtraction is
\begin{equation}
\bar{v}_0^{\alpha}
=
\sum_{pq}
v_{pq}^{\alpha}
(\mathbf P_T)_{pq}
\end{equation}
We can then define
the importance function (in hybrid form\citep{purwanto_pressure_bound}) as
\begin{align}\nonumber
I_{k}(\tau,\Delta\tau, \{\mathbf x_i\}_{i=1}^{k}) 
    =& S_{k}(\tau, \Delta\tau,\{\mathbf x_i - \bar{\mathbf x}_i\}_{i=1}^{k}) \\
 & \times    e^{\mathbf{x}_k\cdot\mathbf{\bar{x}}_k-\frac{\mathbf{\bar{x}}_k\cdot\mathbf{\bar{x}}_k}{2}}
 \label{eq:import}
\end{align}
The phaseless approximation ensures the reality and positivity of walkers using 
a modified importance function,
\begin{align} \nonumber
I_{\text{ph}, k}(\tau,\Delta\tau, \{\mathbf x_i\}_{i=1}^{k}) 
= &
|
I_{k}(\tau,\Delta\tau, \{\mathbf x_i\}_{i=1}^{k}) 
|  \\ 
& \times
\text{max}(0, \cos\theta_k)
 \label{eq:phimport}
\end{align}
where
\begin{equation}
\theta_k
= \text{arg}\left(
S_{k}(\tau,\Delta\tau,\{\mathbf x_i\}_{i=1}^{k}) 
\right).
\label{eq:theta}
\end{equation}
Using these, the $n$-th walker weight at $\tau = k\Delta\tau$ is updated via
\begin{equation}
w_n(\tau+\Delta\tau) = 
I_{\text{ph}, k}(\tau,\Delta\tau, \{\mathbf x_i\}_{i=1}^{k}) 
\times w_n(\tau)
\end{equation}
This completes the description
of the ph-AFQMC algorithm at finite temperature.
\subsubsection{The $T\rightarrow 0$ limit}\label{sec:ztlim}
Converging ph-FT-AFQMC to the
ph-ZT-AFQMC limit
as decreasing $T$
is highly desirable.
This is due to
the remarkable accuracy
of ph-ZT-AFQMC in a variety of systems benchmarked to date.\cite{simons_hubbard_2d,motta_hydrogen,williams_transition_metal_simons,lee2020performance,lee2020utilizing}
Reaching this zero temperature limit
would naturally suggest that
the
ph-FT-AFQMC algorithm is 
accurate at low-temperature.
Unfortunately, reaching the zero temperature limit
is,  in fact, difficult due to the differences in the phaseless constraint between two algorithms as we shall see below.

We first define the zero temperature overlap ratio at $\tau = k\Delta\tau$:\cite{zhang_phaseless}
\begin{align}\nonumber
&S_{k}^{T=0}(\tau,\Delta\tau,\{\mathbf x_i\}_{i=1}^{k})\\
& = \frac
{\langle \psi_T (N)| \hat{B}(\Delta\tau, \mathbf x_k, 0) |\psi(\tau,\{\mathbf x_i\}_{i=1}^{k-1},N) \rangle}
{\langle \psi_T (N)|  \psi(\tau,\{\mathbf x_i\}_{i=1}^{k-1},N)\rangle}
\label{eq:ztoratio}
\end{align}
where $N$ is the number of particles, 
$|\psi_T(N)\rangle$ is the trial wavefunction defined as
\begin{equation}
|\psi_T(N)\rangle
=
\lim_{n\rightarrow\infty}(\hat{B}_T(\Delta\tau,0))^{n} | \psi_0(N)\rangle
\label{eq:psiT}
\end{equation}
with $| \psi_0(N)\rangle$ being some initial wavefunction
and
\begin{equation}
|  \psi(\tau,\{\mathbf x_i\}_{i=1}^{k-1},N)\rangle
=
\prod_{i=1}^{k-1}\hat{B}(\Delta\tau,\mathbf x_i,0) | \psi_T(N)\rangle.
\end{equation}
We note that
we set $\mu=\mu_T=0$ 
because
the ph-ZT-AFQMC
algorithm typically works
in a fixed particle number space.
In other words, the number of particles in $|\psi_T(N)\rangle$ and $|\psi(N)\rangle$
is
$N$ as specified. Therefore, there is no need to have chemical potentials.
This is then used to construct the phaseless importance function in \cref{eq:phimport}, 
which defines the phaseless approximation at $T=0$.

The correspondence between 
the $T\rightarrow0$ limit of ph-FT-AFQMC
and ph-ZT-AFQMC
can be understood by comparing \cref{eq:ftoratio2} and \cref{eq:ztoratio}.
It is then ultimately equivalent to showing
$
\tr(\hat{A}(\tau,\{\mathbf x_i\}_{i=1}^{k-1}))
=
\langle \psi_T (N) |  \psi(\tau,\{\mathbf x_i\}_{i=1}^{k-1},N)\rangle
$ where $\hat{A}$ is defined through \cref{eq:Abrenda}.
In the limit of $T\rightarrow 0$, the number of time slices $n$ becomes $\infty$.
First, we consider
the propagation in the middle of a path.
Namely, we can
assume $1\ll k\ll n$ in this case. One starts from
\begin{align} \nonumber
&\lim_{T\rightarrow0}\tr(\hat{A}(\tau,\{\mathbf x_i\}_{i=1}^{k-1}))\\
&=
\lim_{n\rightarrow\infty}
\sum_M \sum_\alpha
\langle \psi_\alpha(M)|
(\hat{B}_T(\Delta\tau,\mu_T))^{n-k+1} \nonumber \\
&~~~~~~~~~~
\times
\prod_{i=1}^{k-1}\hat{B}(\Delta\tau,\mathbf x_i,\mu)
|\psi_\alpha(M)
\rangle
\end{align}
where the summation over $M$ is done
over states in different particle number sectors and
the summation over $\alpha$ can be thought of as
summing over all possible states in the basis of $|\psi_T(M)\rangle$.
Assuming that $\mu_T$ and $\mu$ are chosen so that it should pick out the $N$-particle sector,
the action of $(\hat{B}_T(\Delta\tau, \mu_T))^{n-k+1}$ to the left when $n\rightarrow\infty$ (as well as $1 \ll k \ll n$)
yields only one term out of the summation. That is, using \cref{eq:psiT}, 
\begin{align}\nonumber
\tr(\hat{A}(\tau,\{\mathbf x_i\}_{i=1}^{k-1}))
&\rightarrow
\langle \psi_T(N)|
\prod_{i=1}^{k-1}\hat{B}(\Delta\tau,\mathbf x_i, \mu)
|\psi_T(N)\rangle\\
&=
\langle \psi_T(N)|
\psi(\tau,\{\mathbf x_i\}_{i=1}^{k-1}, N)\rangle
\label{eq:ztlim}
\end{align}
This then shows that 
the phaseless constraints
are equivalent between
ph-ZT-AFQMC and ph-FT-AFQMC algorithms
when $1 \ll k \ll n$. 

Subtleties arise
when $1\ll k \ll n$ does not hold.
In the ph-FT-AFQMC algorithm,
such a case always happens
towards the completion of a path.
As an extreme case, let us consider
the phaseless constraint at the last imaginary time step in the ph-FT-AFQMC algorithm.
Namely, let us set $k = n$.
With the same assumptions about $\mu$ and $\mu_T$, we have
\begin{align} \nonumber
&\lim_{T\rightarrow0}\tr(\hat{A}(\tau,\{\mathbf x_i\}_{i=1}^{n-1}))\\  \nonumber
&=
\lim_{n\rightarrow\infty}
\sum_\alpha
\langle \psi_\alpha(N)|
\hat{B}_T (\Delta\tau, \mu_T) \\ 
&~~~~~~~~~~\times \prod_{i=1}^{n-1}\hat{B}(\Delta\tau,\mathbf x_i, \mu)
|\psi_\alpha(N)
\rangle
\label{eq:ztissue}
\end{align}
Not only does the summation over $\alpha$ not easily truncate,
but also
this limit no longer corresponds to
the zero temperature limit unlike in \cref{eq:ztlim}.
This is indeed
why one
should not expect 
the ph-FT-AFQMC energy
to approach
the ph-ZT-AFQMC energy in general.
\insertrev{Because of these issues, even the importance function may need to be modified while the same importance function was advocated in ref. \citenum{LiuFTAI2018}. Further investigation on this in the future will be interesting.}

Lastly, there may be an additional complication 
when our assumptions about $\mu$ and $\mu_T$ do not hold exactly.
In other words, it is possible
to have
fluctuation in the number of particles along an imaginary path.
Such fluctuation is not simply due to the fact that we are working
in the grand canonical ensemble.
Instead, it is due to the fact that 
the chemical potential used in $\hat{B}_T$ (i.e., $\mu_T$) is
not necessarily the same as that of the many-body chemical potential used in $\hat{B}$.
Therefore, on average, at a given imaginary time $\tau$,
$\mathbf A$ may not yield the same number of particles as our target $N$.
However, this
is generally only secondary compared to the issue discussed in \cref{eq:ztissue} 
(i.e., the issue of not projecting to $\alpha = 0$)
unless
the underlying system has a near-zero gap.
When the system is metallic,
the number of particle is very sensitive to both chemical potentials in $\hat{B}_T$ and $\hat{B}$.
Therefore, along an imaginary propagation,
the number particle keeps changing even at very low temperature.
Such a subtlety arises in other methods that work in the grand canonical ensemble
such as  low-order perturbation theory, which has been a subject
of active research for some time.\cite{Jha2020}
Nevertheless,
when the gap is not so small,
in the limit of $T\rightarrow 0$,
the number of particles changes
very little as a function of $\mu$.
This makes the effect of particle fluctuation very small.
In principle one can work directly in the canonical ensemble
which removes the need for
the assumptions about $\mu_T$ and $\mu$.\cite{Shen2020}
Nonetheless,
the illustration of \cref{eq:ztissue} still holds
and some modification to the constraint is necessary in the context as well.

\insertrev{These simple illustrations suggest that
 one may impose the phaseless constraint for $T>0$ based on 
 a modified overlap ratio (at $\tau = k\Delta\tau$), (with the same assumptions about $\mu_T$ and $\mu$),
 \begin{equation}
 \tilde{S}_{k}(\tau,\Delta\tau,\{\mathbf x_i\}_{i=1}^{k}) 
 = \frac
 {\tr((\hat{B}_T(\Delta\tau, \mu_T))^{n} \prod_{i=1}^{k}\hat{B}(\Delta\tau,\mathbf x_i,\mu))}
 {\tr((\hat{B}_T(\Delta\tau, \mu_T))^{n} \prod_{i=1}^{k-1}\hat{B}(\Delta\tau,\mathbf x_i,\mu))}
 \label{eq:ztfix}
 \end{equation}
 where 
 the trial density matrix 
 is always
 multiplied up to the $n$-th power
 so that even 
 in the last imaginary-time step 
 one recovers the zero temperature limit properly.
 However, the overall temperature in \cref{eq:ztfix} is always lower than the physical temperature, which may break down in higher temperature regimes.
 Moreover, 
 Examining this constraint would be an interesting research topic in the future,
 but for the purpose of this work we report numerical results 
 based on the constraint in \cref{eq:ftoratio}.}


%

\subsection{Numerical Stabilization}\label{sec:stab}
\insertnew{It is well known that the standard determinant QMC (DQMC) algorithm suffers from numerical instabilities resulting from the repeated multiplication of the $\mathbf{B}$ matrices.\citep{white_dqmc}
This issue can be overcome using the stratification method\cite{tomas2012advancing} which we will now briefly describe.}
We first write a $\mathbf A$ matrix via a column-pivoted QR decomposition (QRCP),
\begin{equation}
\mathbf A
=
\mathbf Q \mathbf R \mathbf \Pi
\end{equation}
where $\mathbf Q$ is an orthogonal matrix, $\mathbf R$ is an upper-triangular matrix, and $\mathbf \Pi$ is a permutation matrix.
We then define 
\begin{equation}
\mathbf A = \mathbf Q \mathbf D \mathbf T
\end{equation}
where 
\begin{align}
\mathbf D &= \text{diag}(\mathbf R) \\
\mathbf T &= \mathbf D^{-1} \mathbf R \mathbf \Pi
\end{align}
The QRCP decomposition is more expensive than matrix multiplications so 
we perform the decomposition once in a while and this frequency is controlled by 
a ``stack'' size parameter $L$. The number of stacks is then specified by $n_\text{stack} = \beta / L$.
The stack size $L$ is set such that one can perform the product of $\mathbf B$ $L/\Delta\tau$ times without numerical instability.

At an imaginary time $\tau$, we are interested in computing
\begin{equation}
\mathbf A (\tau) = \prod_{i = 1}^{n_\text{stack}} \mathbf{A}_i
\end{equation}
The product of stack $\{\mathbf A_i\}$ needs to be performed following the stratification algorithm:

\begin{algorithmic}
\State Compute the QRCP $\mathbf A_1 = \mathbf Q_1 \mathbf D_1 \mathbf T_1$.
\For ($\:\: 2\leq i\leq n_\text{stack}$)
\State Compute $\mathbf C_i = (\mathbf A_i \mathbf Q_{i-1})D_{i-1}$.
\State Compute the QRCP $\mathbf C_i = \mathbf Q_i \mathbf R_i \mathbf \Pi_i$.
\State Set $\mathbf D_i = \text{diag}(\mathbf R_i)$.
\State Compute $\mathbf  T_i = (\mathbf D_i^{-1}\mathbf R_i)(\mathbf \Pi_i \mathbf  T_{i-1})$.
\EndFor
\end{algorithmic}
In the end of this algorithm, we achieve
\begin{equation}
\mathbf A (\tau) = \mathbf Q \mathbf D \mathbf T
\label{eq:qdt}
\end{equation}
Finally, the Green's function is also computed via a numerically stable form,
\begin{equation}
\mathbf G = (\mathbf T^{-T}\mathbf Q^T \mathbf D_b + \mathbf D_s)^{-T} \mathbf D_b \mathbf Q^T
\end{equation}
where
\begin{equation}
D_b(i,i) = 
\begin{cases}
1/|D(i,i)| & \text{if} |D(i,i)| > 1 \\
1 & \text{otherwise}
\end{cases}
\end{equation}
and
\begin{equation}
D_s(i,i) = 
\begin{cases}
D(i,i) & \text{if} |D(i,i)| \leq 1 \\
\text{sgn}(D(i,i)) & \text{otherwise}
\end{cases}
\end{equation}
This can be derived from
\begin{align}
\mathbf G^{-1}
&=
\mathbf I
+\mathbf A \\
&=
\mathbf I + \mathbf Q \mathbf D \mathbf T\\
&=
\mathbf Q
(\mathbf Q^{-1}\mathbf T^{-1} + \mathbf D)
\mathbf T \\
&=
\mathbf Q
\mathbf D_b^{-1}
(\mathbf D_b \mathbf Q^{-1}\mathbf T^{-1} + \mathbf D_s)
\mathbf T
\end{align}
\subsection{Exploiting the stack structure }
It is possible to reuse the QDT factorization of all stacks but one which is the central stack we are propagating.
We write $\mathbf A(\tau)$ as a product of two matrices (left and right blocks):
\begin{equation}
\mathbf A (\tau)
=
\mathbf A_L (\tau)
\mathbf A_R (\tau)
\end{equation}
We note that the QDT factorizations of left and right blocks are already given from the previous propagation.
Namely, we already have
\begin{align}
\mathbf A_L
&=
\mathbf Q_L \mathbf D_L \mathbf T_L\\
\mathbf A_R
&=
\mathbf Q_R \mathbf D_R \mathbf T_R
\end{align}
Then, the numerically stable formation of $\mathbf A(\tau)$ requires only two QRCP calls (as opposed to $n_\text{stack}$ calls described previously)
and also far less matrix-matrix multiplications to do. The algorithm for propagating one time step is
as follows:
\begin{algorithmic}
\State Compute $\mathbf T_L = \mathbf T_L \mathbf B_T^{-1}$.
\State Compute $\mathbf C_{CR} = (\mathbf B \mathbf Q_{R})D_{R}$.
\State Compute the QRCP $\mathbf C_{CR} = \mathbf Q_{CR} \mathbf R_{CR} \mathbf \Pi_{CR}$.
\State Set $\mathbf D_{CR} = \text{diag}(\mathbf R_{CR})$.
\State Compute $\mathbf  T_{CR} = (\mathbf D_{CR}^{-1}\mathbf R_{CR})(\mathbf \Pi_{CR} \mathbf  T_{R})$.
\State Compute $\mathbf C_{LCR} = \mathbf Q_L \mathbf D_L \mathbf T_L \mathbf Q_{CR} \mathbf D_{CR}$.
\State Compute the QRCP $\mathbf C_{LCR} = \mathbf Q_{LCR} \mathbf R_{LCR} \mathbf \Pi_{LCR}$.
\State Set $\mathbf D_{LCR} = \text{diag}(\mathbf R_{LCR})$.
\State Compute $\mathbf  T_{LCR} = (\mathbf D_{LCR}^{-1}\mathbf R_{LCR})(\mathbf \Pi_{LCR} \mathbf  T_{CR})$.
\end{algorithmic}
We therefore achieve
\begin{equation}
\mathbf A = \mathbf Q_{LCR} \mathbf D_{LCR} \mathbf T_{LCR}
\end{equation}
as in \cref{eq:qdt}. The Green's function can then be computed as before.
\subsection{Exploiting the low-rank structure }\label{sec:lr}
He an coworkers\citep{HeLowRank2019} found that at low temperature both $\mathbf A_L$ and $\mathbf A_R$ are low-rank which can enable significant savings ($\mathcal{O}(M/N)$). 
Such low-rank structures are reflected on $\mathbf D_L$ and $\mathbf D_R$ where with a certain threshold, we can
approximate them as
\begin{align}
\mathbf D_L &\approx \mathbf d_L\\
\mathbf D_R &\approx \mathbf d_R
\end{align}
where
\begin{equation}
\mathbf d_{L/R} (i,i) = 
\begin{cases}
D_{L/R}(i,i) & \text{if} {\: |D_{L/R}(i,i)| \ge}\: \text{threshold} \\
0 & \text{otherwise}
\end{cases}
\label{eq:thresh}
\end{equation}
We denote the rank of $\mathbf d_{L/R}$ to be $m_{L/R}$ and show how scaling reduction can be achieved in terms of these ranks.
From now on, we will only work in the reduced dimension provided by $\mathbf d_{L/R}$.
$\mathbf d_{L/R}$ is a $m_{L/R} \times m_{L/R}$ matrix that is much smaller than the original $M \times M$ matrix.

We write
\begin{align}
\mathbf A_L
&=
\mathbf q_L \mathbf d_L \mathbf t_L\\
\mathbf A_R
&=
\mathbf q_R \mathbf d_R \mathbf t_R
\end{align}
where $\mathbf q_{L/R}$ is a matrix of dimension $M\times m_{L/R}$  and $\mathbf t_{L/R}$ is a matrix of dimension $m_{L/R} \times M$.
To maximize cost saving, one needs to modify the stratification algorithm further.
The most efficient propagation with stratification can be done as follows (assuming we sampled $\mathbf B$):
\begin{algorithmic}
\State Compute
$\mathbf t_L = \mathbf t_L \mathbf B_T^{-1}$ 
\State Compute
$\mathbf c_{CR}
= \mathbf B \mathbf q_R \mathbf d_R$ 
\State Compute
$\mathbf c_{LCR}
= \mathbf d_L \mathbf t_L \mathbf c_{CR}$
\State Compute the QRCP $\mathbf c_{LCR} = \mathbf q_{LCR} \mathbf r_{LCR} \mathbf \pi_{LCR}$
($\mathbf q_{LCR}$ is $m_L \times m_T$ and $\mathbf r_{LCR}$ is $m_T\times m_R$ where $m_T = \text{min}(m_R,m_L)$)
\State Compute $\mathbf q_{LCR} = \mathbf q_{L}\mathbf q_{LCR}$. 
\State Set $\mathbf d_{LCR} = \text{diag}(\mathbf r_{LCR})$.
\State Compute $\mathbf  t_{LCR} = (\mathbf d_{LCR}^{-1}\mathbf r_{LCR})(\mathbf \pi_{LCR} \mathbf  t_{R})$.
\end{algorithmic}
We then achieve
\begin{equation}
\mathbf A = \mathbf q \mathbf d \mathbf t
\end{equation}
where $\mathbf q$ is an $M\times m_T$ matrix, $\mathbf d$ is an $m_T\times m_T$ diagonal matrix, and $\mathbf t$ is a $m_T\times M$ matrix.
Using these reduced dimension matrices and the Woodbury identity, we can compute the Green's function at a reduced cost:
\begin{align}
\mathbf G &= (\mathbf I + \mathbf q \mathbf d \mathbf t)^{-1} \\
& = 
\mathbf I - \mathbf q(\mathbf d^{-1} + \mathbf t \mathbf q)^{-1}\mathbf t
\end{align}
where $(\mathbf d^{-1} + \mathbf t \mathbf q)^{-1}$ occurs in the dimension of $m_T\times m_T$ and
other matrix multiplications are done at the cost of $\mathcal O(M^2 m_T)$.
Similarly, using the matrix determinant lemma
\begin{equation}
\det(\mathbf I_M + \mathbf q \mathbf d \mathbf t)
=
\det(\mathbf I_{m_T} + \mathbf d \mathbf t \mathbf q )
\end{equation}
The determinant evaluation occurs in the dimension of $m_T \times m_T$ as well.

Furthermore, we note that the evaluation of $\mathbf I_{m_T} + \mathbf t \mathbf q \mathbf d$ needs to be done by the usual stabilization algorithm:
\begin{align}
\mathbf I_{m_T} + \mathbf d \mathbf t \mathbf q
&=
(\mathbf q^{-1}\mathbf t^{-1} + \mathbf d)
\mathbf t \mathbf q \\
&=
(\mathbf q^{-1}\mathbf t^{-1} \mathbf d_b + \mathbf d_s)
\mathbf d_b^{-1}
\mathbf t \mathbf q 
\end{align}
From this, we evaluate the determinant as well as the Green's function.

For measurements, we use 1RDM, $\mathbf P$, which is now expressed as
\begin{equation}
\mathbf P = \mathbf t^T(\mathbf d^{-1}+\mathbf t \mathbf q)^{-T} \mathbf q^T
\end{equation}
This $\mathbf P$ is also of low-rank and this structure can be exploited to accelerate the local energy evaluation and other measurements.
\subsection{Uniform Electron Gas}
The uniform electron gas (UEG) model is usually defined and solved in the plane-wave basis set. We will follow this convention in this work as well.
The kinetic energy operator is defined as
\begin{equation}
\hat{T} = \sum_{\mathbf K} \frac{|\mathbf K|^2}{2} a_{\mathbf K}^\dagger a_{\mathbf K}
\end{equation}
where $\mathbf K$ here is a plane-wave vector.
The electron-electron interaction operator is (in a spin-orbital basis)
\begin{equation}
\hat{V}_\text{ee} = \frac{1}{2\Omega	}
\sum_{\mathbf{K}\ne\mathbf 0,\mathbf{K}_1,\mathbf{K}_2}
\frac{4\pi}{|\mathbf{K}|^2}
a_{\mathbf{K}_1+\mathbf{K}}^\dagger
a_{\mathbf{K}_2-\mathbf{K}}^\dagger
a_{\mathbf{K}_2}
a_{\mathbf{K}_1}
\end{equation}
where $\Omega$ is the volume of the unit cell. 
Lastly, the Madelung energy $E_M$ should be included to account for the self-interaction of the Ewald sum under periodic boundary conditions and\citep{schoof_prl}
\begin{equation}
E_M = -2.837297 \times \left(\frac{3}{4\pi}\right)^{1/3}N^{2/3}r_s^{-1},
\end{equation}
where $N$ is the number of electrons in the unit cell and $r_s$ is the Wigner-Seitz radius.
We define the UEG Hamiltonian as a sum of these three terms,
\begin{equation}
\hat{H} = \hat{T} + \hat{V}_\text{ee} + E_M
\end{equation}
The two-body Hamiltonian $\hat{V}_\text{ee}$ needs to be rewritten as a sum of squares to employ
the AFQMC algorithm. It was shown in Ref. \citenum{afqmc_pwpsp} that
\begin{equation}
\hat{V}_\text{ee} = \frac14 \sum_{\mathbf Q \ne \mathbf 0}
\left[
\hat{A}^2(\mathbf Q)
+
\hat{B}^2(\mathbf Q)
\right]
\end{equation}
where
\begin{equation}
\hat{A}(\mathbf Q) = \sqrt{\frac{2\pi}{\Omega |\mathbf Q|^2}}
\left(
\hat{\rho}(\mathbf Q)
+ \hat{\rho}^\dagger(\mathbf Q)
\right)
\end{equation}
and
\begin{equation}
\hat{B}(\mathbf Q) = i\sqrt{\frac{2\pi}{\Omega |\mathbf Q|^2}}
\left(
\hat{\rho}(\mathbf Q)
- \hat{\rho}^\dagger(\mathbf Q)
\right)
\end{equation}
with a momentum transfer operator $\hat{\rho}$ defined as
\begin{equation}
\hat{\rho}(\mathbf Q) = \sum_\sigma \sum_{\mathbf K} a^\dagger_{\mathbf K + \mathbf Q, \sigma}
a^\dagger_{\mathbf K, \sigma} \Theta\left(E_\text{cut} - \frac{|\mathbf K + \mathbf Q|^2}{2}\right)
\end{equation}
where
$\Theta$ is the Heaviside step function and $E_\text{cut}$ is the kinetic energy cutoff in the simulation.
The HS operators $\hat{\mathbf v}$ are now $\hat{A}(\mathbf Q)$ and $\hat{B}(\mathbf Q)$.

The local energy $\epsilon_n(\tau)$ for the UEG then reads
\begin{align}\nonumber
\epsilon_n(\tau)
&=
E_M + E_1 + E_2
\end{align}
where 
the one-body energy, $E_1$, is
\begin{equation}
E_1 = \sum_{\mathbf K} \frac{|\mathbf K|^2}{2} P_{\mathbf K \mathbf K}
\end{equation}
and
the two-body energy, $E_2$, is
\begin{equation}
E_2 = \frac{1}{2\Omega}\sum_{\mathbf{Q}\ne\mathbf 0}
\frac{4\pi}
{|\mathbf{Q}|^2} \left(
\Gamma_{\mathbf Q}
-
\Lambda_{\mathbf{Q}}\right)
\end{equation}
with
the Coulomb two-body density matrix $\Gamma_{\mathbf Q}$,
\begin{equation}
\Gamma_{\mathbf Q} = 
\left(\sum_{\mathbf K_1}P_{\mathbf{K}_1+\mathbf{Q},\mathbf{G}_1}\right)
\left(\sum_{\mathbf K_2}P_{\mathbf{K}_2-\mathbf{Q},\mathbf{G}_2}\right)
\end{equation}
and the exchange two-body density matrix $\Lambda_{\mathbf Q}$,
\begin{equation}
\Lambda_{\mathbf{Q}} = 
\sum_{\mathbf{K}_1\mathbf{K}_2}
P_{\mathbf{K}_1+\mathbf{Q},\mathbf{K}_2}
P_{\mathbf{K}_2-\mathbf{Q},\mathbf{K}_1}
\end{equation}
The formation of $\Gamma_{\mathbf Q}$ costs $\mathcal O(M^2)$ whereas the formation of $\Lambda_{\mathbf Q}$ takes $\mathcal O(M^3)$ amount of work.
Therefore, the evaluation of the exchange contribution is the bottleneck in the local energy evaluation.
As noted in Ref.~\citenum{Zhang2008} the evaluation of the energy (and propagation) can be accelerated using fast Fourier transforms,  
we found this to be slower 
than an algorithm
based on sparse linear algebra
for system sizes considered in this work.
Therefore, our implementation 
is exclusively
based on sparsity.

Unlike the ph-ZT-AFQMC algorithm,
finite temperature estimators are
all ``pure'' as opposed to ``mixed'',
which 
does not require any special
treatments
to measure
expectation values of operators that do not commute with $\hat{H}$.
For instance, the one-body and two-body energies can be
straightforwardly read off from the local energy evaluation.

\section{Computational Details}
All simulations were performed with the open source \texttt{PAUXY} code.\citep{pauxy} \texttt{HANDE} was used to perform FCI calculations.\citep{hande_jctc,HANDEdev}
Q-Chem was used to crosscheck some of the mean-field finite temperature calculations.\cite{Shao2015}

We used 640 walkers along with the `comb' population control method.\citep{BoothComb2009,NguyenCPMCLab2014} 
For $\Theta \le 0.25$ we found that the comb algorithm introduced significant population control biases so we switched here to the pair branch algorithm.\citep{wagner_qwalk}
Multiple time steps were employed
depending on $r_s$, which are $\Delta\tau=0.05$ for $r_s \le 1$, $\Delta\tau = 0.025$ for $r_s = 2$, and $\Delta\tau=0.005$ for  $r_s>2$ all in reduced units. \insertnew{This choice was made to maintain the time step to be roughly the same throughout all $r_s$ values in the atomic unit.}
Given our time step error and population control bias study in the 54-electron super cell at zero temperature,\cite{LeeUEG2019} we expect that the time step and population control biases will not affect the conclusions of our study.

We typically averaged results over between 50-100 independent simulations, which typically were run with 160 CPUs for 10 hours.
We used a free electron trial density matrix and tuned the chemical potential $\mu_T$ such that $\langle \hat{N}\rangle_T = \bar{N}$.
The interacting chemical potential was then tuned to also match $\bar{N}$.
To determine the interacting chemical potential we first use root bracketing based off a single step of the ph-FT-AFQMC algorithm (i.e. without considering error bars) until the chemical potential was determined to about an accuracy of 1\%, which we call $\tilde{\mu}$.
Following this we perform 5 poorly converged simulations (320 walkers averaged over 10 simulations) in the interval of $[0.975\tilde{\mu},1.025\tilde{\mu}]$.
We next perform a weighted least squares fit to these 5 data points to determine the optimal chemical potential $\mu^*$.
Finally, we ran at this $\mu^*$ using a larger walker number and average over a large number of independent simulations to get the final results reported here.
We find the procedure works most of the time at the lower temperatures considered here, although some care needs to be taken at higher temperatures where the electron number varies more significantly with $\mu$.
Simulation input and output is available at Ref.\citenum{zenodo}.
We will primarily focus on the regime of $0.5 \le r_s \le 4$ and $\Theta \le 1$, i.e. the warm dense regime, where $\Theta=T/T_F$ and $T_F$ is the Fermi temperature of an unpolarized UEG. 
Box sizes were fixed via $L = \left(\frac{4}{3}\pi \bar{N}\right)^{1/3} r_s$ to enable comparison to results in the canonical ensemble.
\insertnew{We note that
the reduced unit system depends on the value of $r_s$ because the Fermi temperature 
depends on $r_s$ as
\begin{equation}
k_B T_F = \frac12\left(\frac{9\pi}{4}\right)^{2/3} \frac{1}{r_s^2}
\label{eq:fermiT}
\end{equation}
where $k_B$ is the Boltzmann constant. Due to this, for a given reduced temperature $\Theta$ we observe the total path length (i.e., $1/(k_BT)$) in atomic unit to increase as $\mathcal O(r_s^2)$, which makes the higher $r_s$ values computationally costly.}
A low-rank threshold of $10^{-6}$ in \cref{eq:thresh} was used in every calculation.
\section{Results}
\subsection{Assessing the Phaseless Constraint}\label{sec:accuracy}
In ph-FT-AFQMC, the main source of error is the bias introduced by the phaseless constraints. The impact of this bias is heavily dependent on the quality of trial density matrices.
Here, we will employ a simple non-interacting one-body trial density matrix based on the kinetic energy operator.
Namely,
\begin{equation}
(\mathbf{B}_T)_{\mathbf K, \mathbf K}
=
e^{-\Delta\tau \left(\frac{|\mathbf K|^2}2 - \mu_T \right)},
\end{equation}
which we call a free-electron (FE) trial density matrix with a trial chemical potential, $\mu_T$.

\subsubsection{A study of $\bar{N} = 2$ with $M=7$}
To assess the quality of ph-FT-AFQMC with FE density matrix we can compare to exact full configuration interaction results (FT-FCI) which is possible in a very small finite basis set.\citep{KouFTFCI2014} We study the UEG model with only 7 planewaves ($M=7$) and 
tune the chemical potential to reach
an on-average 2-electron system.

\begin{figure}
    \centering
    \includegraphics[scale=0.90]{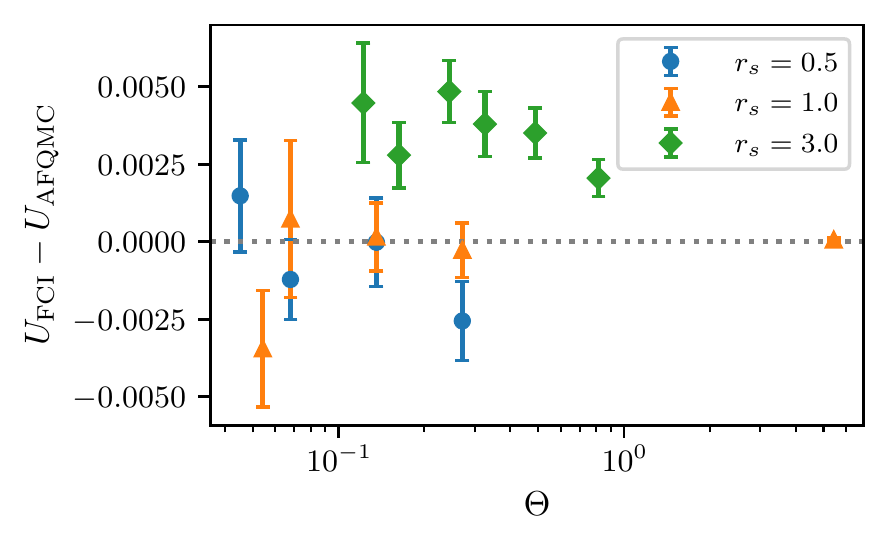}
    \caption{Comparison of ph-FT-AFQMC internal energies ($E_h$) to exact (FCI) results as a function of temperature for $\bar{N}=2$ for different values of $r_s$ and $M=7$.
    }
    \label{fig:fci_comp}
\end{figure}

The results of this comparison are plotted in \cref{fig:fci_comp} where we compare to a small system size ($\bar{N} = 2$) as a function of $r_s$ and $\Theta$.
At first look the figure looks reasonable: ph-FT-AFQMC agrees with FT-FCI very well for low $r_s$ (high densities) which is consistent with results at zero temperature,\citep{LeeUEG2019} and the results disagree more as $r_s$ increases and as the temperature decreases.
\insertnew{Since the $T=0$ ph-ZT-AFQMC results are essentially exact for this system at all $r_s$ values considered here,
this results unfortunately are a manifestation of
the discrepancy between ph-FT-AFQMC and ph-ZT-AFQMC at the zero-temperature limit.}
For example at $r_s=3$ and $T=0$ the ph-ZT-AFQMC energy is -0.23968(3) $E_h/e$ which is statistically identical to the FCI result of -0.23968 $E_h/e$. 
This suggests that the phaseless constraint at finite temperature differs from that at zero temperature, and that it is potentially considerably larger.

\insertnew{In \cref{sec:ztlim}, we analytically showed that two factors (constraint and chemical potential) can make the zero-temperature limit unreachable using the ph-FT-AFQMC algorithm.}
\insertnew{Incorporating other observations made in prior works,\cite{HeSCFT2019} we mention a total of four possible reasons for this disagreement}:
\begin{enumerate}
\item 
{\it the issue of chemical potential mismatch}
Since we work in different ensembles (FT with the grand canonical ensemble and ZT with the canonical ensemble),
it is important to tune $\mu$ and $\mu_T$ properly
so that
the number of electrons does not fluctuate so much 
along an imaginary time path as assumed throughout the discussion in \cref{sec:ztlim}.
Based on \cref{fig:im_time_nav}, 
we suspect that this effect is quite minor
since despite the fact that the average particle number deviates from the desired value along the path the effect of this deviation in other physical observables is negligible.

\begin{figure}
    \centering
    \includegraphics[scale=0.90]{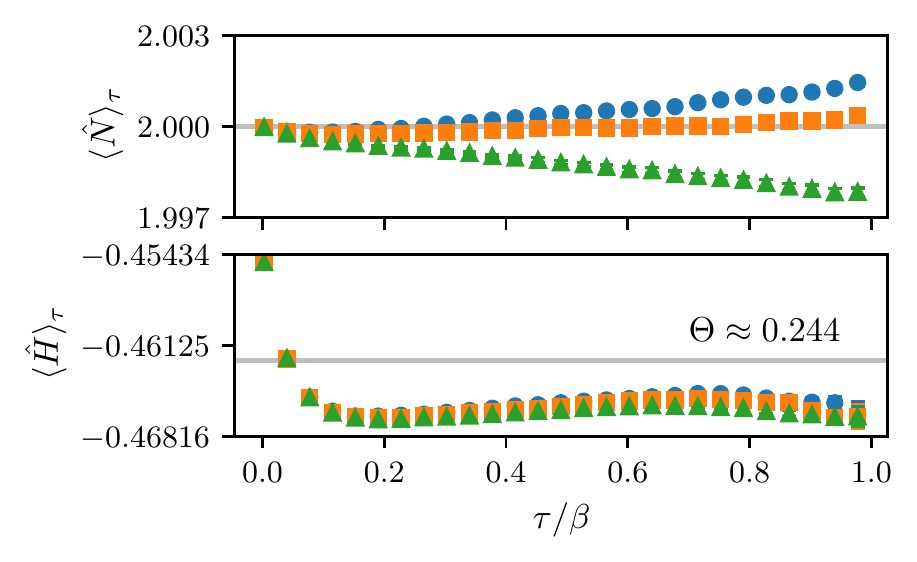}
    \caption{Imaginary time dependence of the average electron number and total energy in ph-FT-AFQMC for the $r_s=3$, $N=2$, $M=7$ UEG model at $\beta = 20$ $E_\mathrm{h}^{-1}$ or $\Theta \approx 0.244$ which is nearly the zero $T$ limit in this small supercell. The three different data sets correspond to 3 chemical potentials chosen around $\mu^*_{\mathrm{FCI}}$. In the lower panel the horizontal line represents the FCI total energy at this density. We used 2048 walkers, the pair-branch population control aglorithm, averaged over 30 independent simulations to obtain the error bars and used a timestep of 0.05 $E_\mathrm{h}^{-1}$.}
    \label{fig:im_time_nav}
\end{figure}
\begin{figure}
    \centering
    \includegraphics{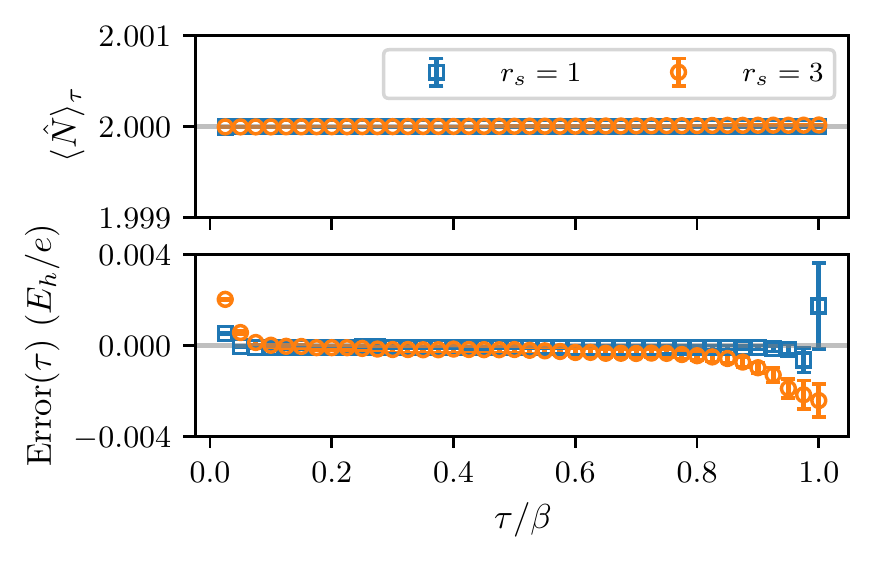}
    \caption{Imaginary time dependence of the average electron number and error in the total energy ($u_{\mathrm{ph-FT-AFQMC}}(\tau)-u_{\mathrm{FCI}})$) in ph-FT-AFQMC for the $N=2$, $M=7$ UEG model for $r_s = 1$ (blue squares) and $r_s=3$ (orange circles). For $r_s=1$ we chose $\beta = 10 $ $E_\mathrm{h}^{-1}$ ($\Theta \approx 0.05$) and for $r_s=3$ we chose $\beta = 40 $ $E_\mathrm{h}^{-1}$ ($\Theta \approx 0.122$). The temperatures were chosen such that they corresponded to the zero temperature limit. We used 2048 walkers, the pair-branch population control algorithm, averaged over 30 independent simulations to obtain the error bars and used a timestep of 0.05 $E_\mathrm{h}^{-1}$.}
    \label{fig:im_time_rs}
\end{figure}
%
%

\item {\it the fundamental differences between the two constraints}
A second possibility is that the constraints are fundamentally different even if $\mu$ and $\mu_T$ are chosen properly.
This is then the direct consequence of the illustration suggested by \cref{eq:ztissue} in \cref{sec:ztlim}.
Indeed as can be seen in \cref{fig:im_time_rs} we see that this appears to be the case.
Towards the end of the path we see that the ph-FT-AFQMC results begin to deviate substantially from the FCI result, with the magnitude of the deviation increasing with $r_s$. This is consistent with the fact that larger $r_s$ requires longer imaginary time to project out the ground state.
This longer projection time results in a larger portion of the path using a determinant ratio that does not resemble the zero temperature overlap ratio.
In the middle of the path results are close to exact and the algorithm more closely resembles ph-ZT-AFQMC.

\item {\it the difference between trial density matrix and trial wavefunction}
A third possibility is that the FE trial density matrix is not appropriate to reproduce the correct zero temperature trial wavefunction used in ph-AFQMC.
This is an important concern in general,\citep{HeSCFT2019} however at the high densities and small systems sizes considered here we do not expect there to be any unrestricted Hartree--Fock solutions. Furthermore, for the UEG, RHF is equivalent to a free electron trial in a closed shell system. 
Nevertheless to verify this, we tested the thermal Hartree--Fock density matrix,\citep{MerminTHF1963} however we see from \cref{fig:thf} this choice makes essentially no practical difference.
\begin{figure}
    \centering
    \includegraphics[width=0.5\textwidth]{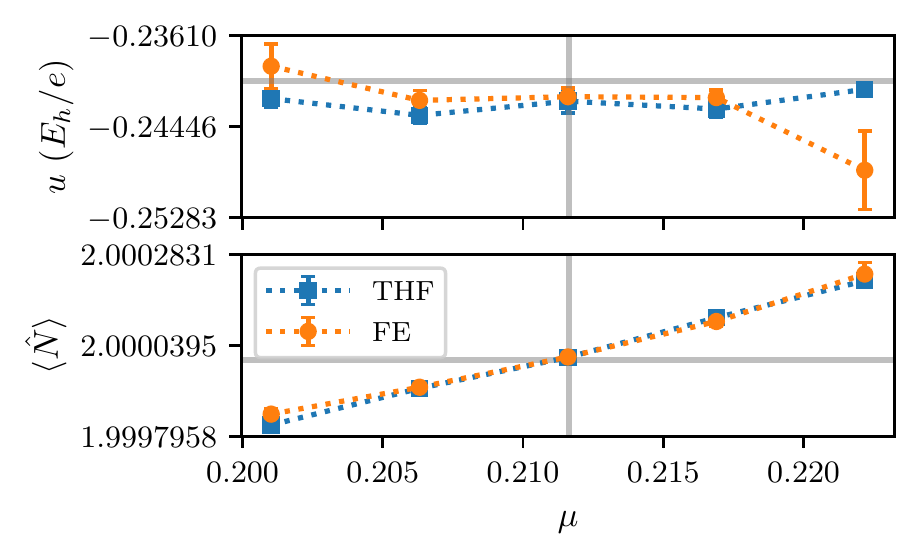}
    \caption{Comparison between the ph-FT-AFQMC internal energy per electron $(u)$ and the average particle number as a function of the chemical potential with a free-electron (FE) and thermal Hartree--Fock (THF) like trial density matrix.
    The vertical (horizontal) lines represent the FCI values for the chemical potential (energy and electron number). The system considered here is  $\bar{N}=2, M=7, r_s = 3$ with $\beta = 40 E_h^{-1}$.}
    \label{fig:thf}
\end{figure}

\item {\it time-reversal symmetry breaking} The final possibility is that as the phaseless constraint breaks imaginary time symmetry in the estimators,\citep{HeSCFT2019} and we should instead average across time slices as suggested in Ref.\citenum{HeSCFT2019}. Interestingly we find that at $r_s=3$ this averaging procedure biases results to be above the exact value (see \cref{fig:averaging}), and again does not resolve this discrepancy with the zero temperature algorithm. This bias can also be understood in the context of \cref{fig:im_time_rs} where the value of the energy along the imaginary time path is not symmetric with respect to $\tau$.
\insertnew{While the time-averaged estimators were found to be more accruate in the Hubbard model,\cite{HeSCFT2019} given our results one should not expect this to be a general solution to this problem.}
\begin{figure}
    \centering
    \includegraphics{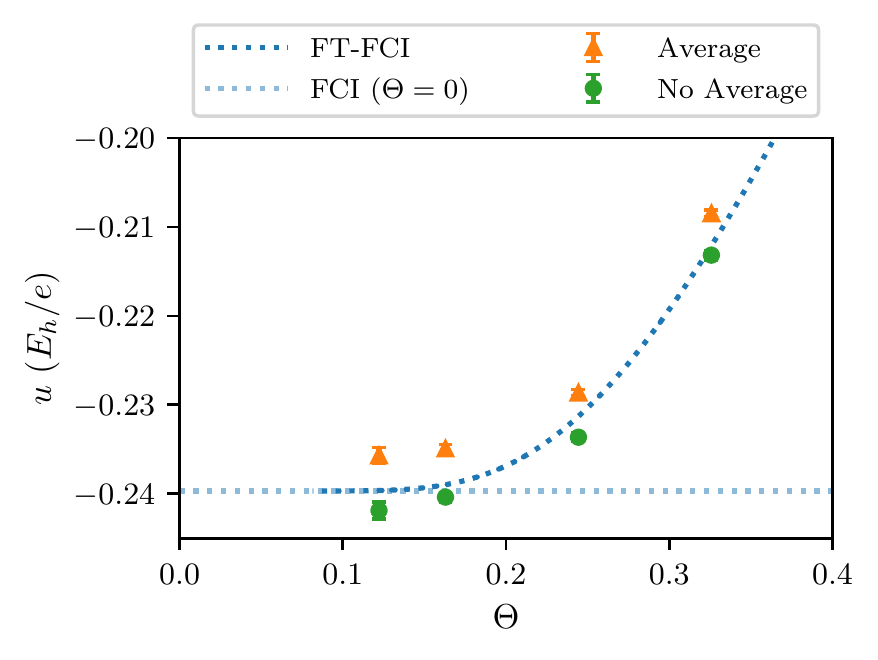}
    \caption{Comparison between time slice averaged energies and the normal asymmetric estimator for the $M=7$, $\bar{N}=2$, $r_s=3$ benchmark system. Error bars where not visible are smaller that the markers.
    \insertrev{``No Average'' refers to calculations where the global energy estimate is computed at $\tau=\beta$ where ``Average'' averages the global estimate across different time slices as suggested in ref. \citenum{HeSCFT2019}.}
    }
    \label{fig:averaging}
\end{figure}
\end{enumerate}

Despite these concerns it is also clear from \cref{fig:fci_comp} that ph-FT-AFQMC performs quite well with only small deviations seen from exact result 
in the regimes we are interested in (i.e., $\Theta < 1$ and $r_s < 3$).
\subsubsection{A study of $\bar{N} = 66$ with $M=57$}
We now study a larger UEG supercell considering $\bar{N} = 66$ with $M=57$.
Obviously, such a parameter set-up would have a very large
basis set incompleteness error and therefore
should not be used to draw any physically meaningful results.
Nevertheless, for this basis set size,
FT-CCSD results are available for several $\Theta$ and $r_s$ values \citep{White2018}
to which we compare ph-FT-AFQMC results.
\insertnew{We use the FE trial density matrix as before since we do not have any UHF solutions at zero-temperature below $r_s=3$ as shown in \cref{sec:symbreak}.}

\begin{figure}
    \centering
    \includegraphics{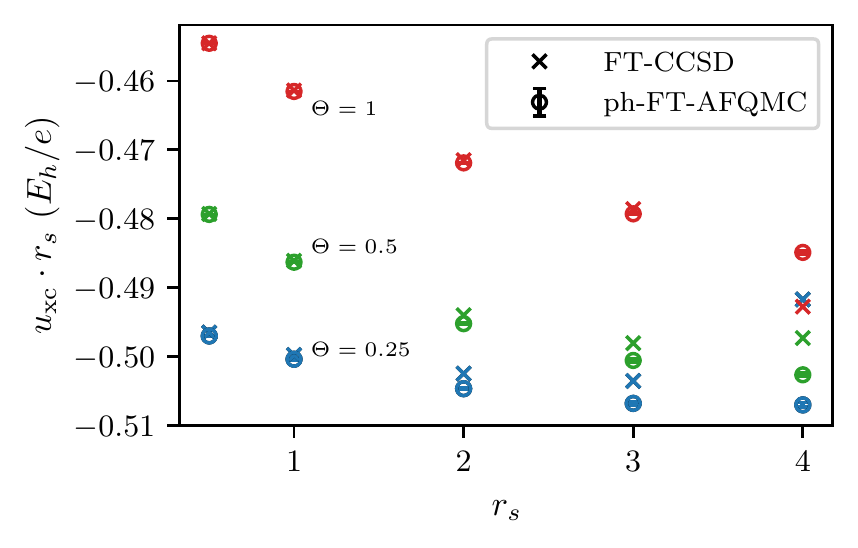}
    \caption{Comparison between ph-FT-AFQMC and FT-CCSD exchange-correlation energies as a function of temperature for $\bar{N}=66$, $M=57$. FT-CCSD energies are reproduced from Ref.\citenum{White2018}.\joonho{Fionn, for $\Theta = 0.25$ I am worried that ph-FT-AFQMC may be biased due to the population control algorithm. It seems that the energy is actually not monotonic going from $r_s = 3$ to $r_s = 4$?}}
    \label{fig:uxc_ccsd}
\end{figure}

The result of this comparison is shown in \cref{fig:uxc_ccsd}.
We see that ph-FT-AFQMC agrees well with FT-CCSD for low $r_s$ in all temperatures considered here.
As we observed in the study of $\bar{N} = 2$,
the performance of ph-FT-AFQMC at low temperature may not be as good as the ph-ZT-AFQMC in the zero temperature limit.
Nonetheless, we found accurate results at low $r_s$ in the case of $\bar{N} = 2$ and
we observe consistently accurate
results at a larger supercell ($\bar{N} = 66$) when compared against FT-CCSD.
CCSD was shown to be accurate for low $r_s$ such as $r_s = 0.5$ in the zero temperature benchmark\cite{neufeld_ccmc} so
this helps us build confidence on the performance of ph-FT-AFQMC at low $r_s$.

However, the quality of FT-CCSD is expected to gradually degrade as $r_s$ increases and $\Theta$ decreases
as indicated by its zero temperature benchmark study.\cite{neufeld_ccmc}
Consequently, at $r_s = 4$ FT-CCSD exchange-correlation energies show erratic behavior
changing the energy trend as a function of $r_s$ completely.
ph-FT-AFQMC, on the other hand, shows a smooth monotonic behavior as a function of $r_s$ in all temperatures.
Given the superior performance of ph-ZT-AFQMC compared to CCSD in the UEG model,\cite{LeeUEG2019}
this result is not surprising.

\subsection{Efficacy of the Low-Rank Truncation}

\begin{figure*}
    \centering
    \includegraphics{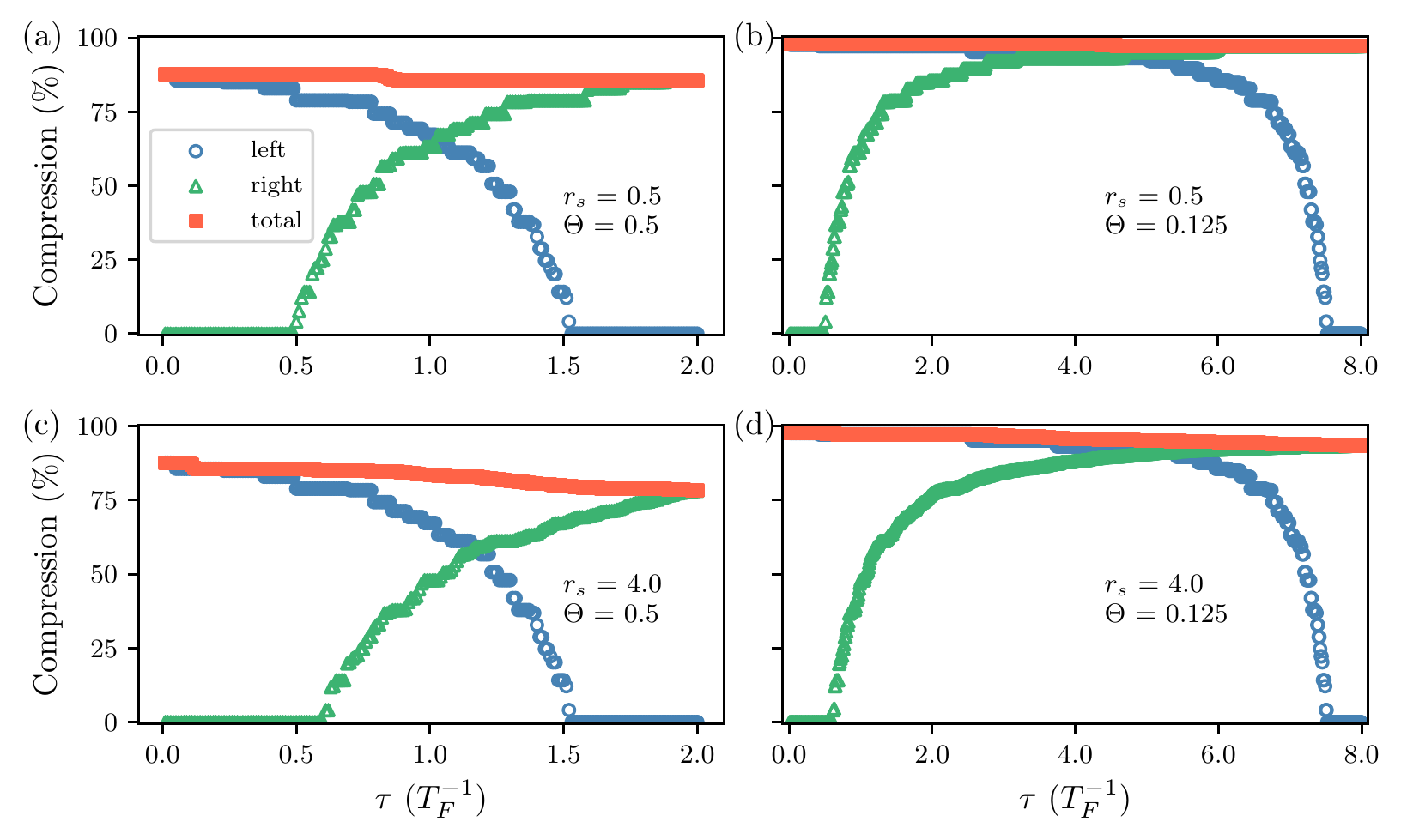}
    \caption{Demonstration of low-rank compression efficiency (\%) in left, right, and total blocks for
    (a) $r_s=0.5$ and $\Theta = 0.5$, 
    (b) $r_s=0.5$ and $\Theta = 0.125$,
    (c) $r_s=4.0$ and $\Theta = 0.5$, and
    (d) $r_s=4.0$ and $\Theta = 0.125$.
    All calculations are done with $M = 1189$ and $\mu$ was chosen so that the one-body trial density matrix satisfies $\bar{N} = 14$.
    A threshold of $10^{-6}$ was used in \cref{eq:thresh}.
     }
    \label{fig:low_rank}
\end{figure*}

In \cref{fig:low_rank}, 
we show the practical effectiveness of the low-rank truncation discussed in \cref{sec:lr}.
We evaluate
the
compression percentage
based on the ratio between the rank
of
left ($m_L$), right ($m_R$), and total ($m_T$) and
the number of basis functions ($M$).
Namely,
\begin{equation}
c_i = 
\left(1 - \frac{m_i}M\right) \times 100 \%
\end{equation}
where $i\in\{L,R,T\}$. 
The set of parameters that we chose to look at this is 
$\bar{N} \simeq 14$, $M = 1189$, $r_s = 0.5, 4.0$, and $\Theta = 0.5,0.125$. 
For the purpose of demonstration,
tuning chemical potential is not 
important so we set $\mu= \mu_T$ which ensures a correct average number of particles in the trial density matrix.

Comparing \cref{fig:low_rank} (a) and (b), we see the
effect of the temperature change at $r_s = 0.5$. 
As expected the lower the temperature, we observe the higher compression ratio.
In fact, we are in the limit where $M \gg \bar{N}$ so the low-rank compression becomes very effective.
As we move along an imaginary path, 
we see that the compression efficiency 
decreases for left (trial density matrix blocks)
and increases for right (sampled blocks).
This is the feature of our propagation algorithm, which
replaces $\hat{B}_T$ by $\hat{B}$ each time step.
We observe nearly constant compression efficiency for the total block.
We obtain about 80\% compression in $m_T$ at $\Theta = 0.5$ and nearly 100\% compression in $m_T$ at $\Theta = 0.125$.
A similar conclusion can be drawn from \cref{fig:low_rank} (c) and (d).
We see only small changes in the compression efficiency when changing from $r_s = 0.5$ to $r_s = 4.0$. 

While this technique is useful in speeding up the calculation in general,
one needs a small $\bar{N}/M$ ratio and low temperature for a large saving.
In our work, among larger calculations,
the biggest saving
was made in the case of
$\bar{N} = 66$
and $M = 485$ at $\Theta = 0.125$, which will be presented below. \insertnew{For this particular example, we did not find the cost saving to be substantial due to its sizable $\bar{N}/M$, but the compression algorithm was still used for the numerical stability.}

\subsection{Comparison to Other Approaches}

In \cref{sec:accuracy},
we focused on
assessing the accuracy of ph-FT-AFQMC   on a very small supercell 
for a given finite basis set.
The results from \cref{sec:accuracy} suggest that ph-FT-AFQMC is quite accurate for $r_s \le 3$ and it is now important to assess 
the utility of this method
in more realistic calculations
specifically towards
the CBS limit.



\begin{figure}
    \centering
    \includegraphics{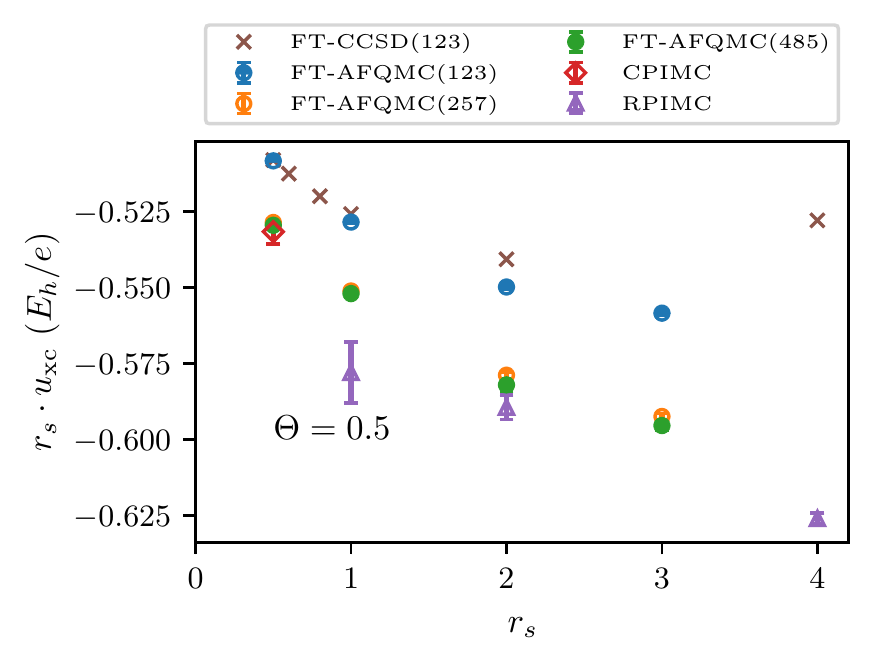}
    \caption{Basis set convergence of the ph-AFQMC exchange-correlation ($u_\text{xc}$) energy at $\Theta=0.5$, $\bar{N}=66$.
    CPIMC results were converged to the basis 
    limit.\citep{DornheimUnpolarized2016}
    \joonho{Check the sample size for $r_s=3,4$ at $M=257$ and remove the data points if the sample size is not enough}
    }
    \label{fig:exc_rs}
\end{figure}

In \cref{fig:exc_rs} we investigate the magnitude of the basis set error as a function of $r_s$ at $\Theta=0.5$ where there exist previous exact CPIMC results as well as RPIMC data.
We see that the ph-FT-AFQMC results are indistinguishable from the CPIMC results at $r_s=0.5$ when for $M\ge 257$ \insertnew{on the plotted scale}.
We also see that $M=485$ is sufficient to converge results up to $r_s=2$ on this scale.
As seen in previous studies we find the RPIMC results to be too low, however the magnitude of this bias seems to decrease with increasing $r_s$.
We can also ascribe the deviation of the FT-CCSD result from the PIMC data to the basis set size.
We see that FT-CCSD and \insertrev{ph-FT-AFQMC} agree well for $M=123$ (the largest basis set considered in Ref.\citenum{White2020}) up to $r_s=1$ with the FT-CCSD results deviating from the expected smooth trend with $r_s$ beyond this.
Unfortunately we find obtaining ph-FT-AFQMC data beyond $r_s=2$ to be too expensive beyond $M=257$ due to the smaller timesteps required (equivalently the larger values of $\beta$ required as $r_s$ increases as in \cref{eq:fermiT}).
Nevertheless the results look promising and it is possible that reliable results could be obtained for even larger $r_s$ values if more computational resources were expended.

With the confidence that the basis set error and phaseless error are under control in this parameter regime we next look towards the thermodynamic limit.
To reach the thermodynamic limit we use finite size corrections calculated from the finite temperature random phase approximation.\citep{vignale_qtel,tanaka1986thermodynamics}
The subject of finite size corrections is treated exhaustively elsewhere,\citep{Groth2016,MaloneThesis2017,dornheim_review} and we will not discuss them in any detail.
In essence, we add a correction $\Delta u_\mathrm{xc}(r_s,\Theta,N)$ to our QMC results where
\begin{equation}
    \Delta u_\mathrm{xc}(r_s,\Theta,N) = u_{\mathrm{xc}}^{\mathrm{RPA}}(r_s,\Theta,\infty)- u_{\mathrm{xc}}^{\mathrm{RPA}}(r_s,\Theta,N), 
\end{equation}
where $u_{\mathrm{xc}}^{\mathrm{RPA}}$ can be computed numerically through derivatives of the exchange-correlation free energy.
Detailed equations and the code necessary for this are available in Refs. \citenum{MaloneThesis2017} and \citenum{uegpy}. 
We also verified the accuracy of these corrections across $r_s$ and $\Theta \ge 1$ independently in our Supplementary Material\citep{supplement} using existing CPIMC and PB-PIMC data.\citep{DornheimUnpolarized2016}

\begin{figure*}
    \centering
    \includegraphics{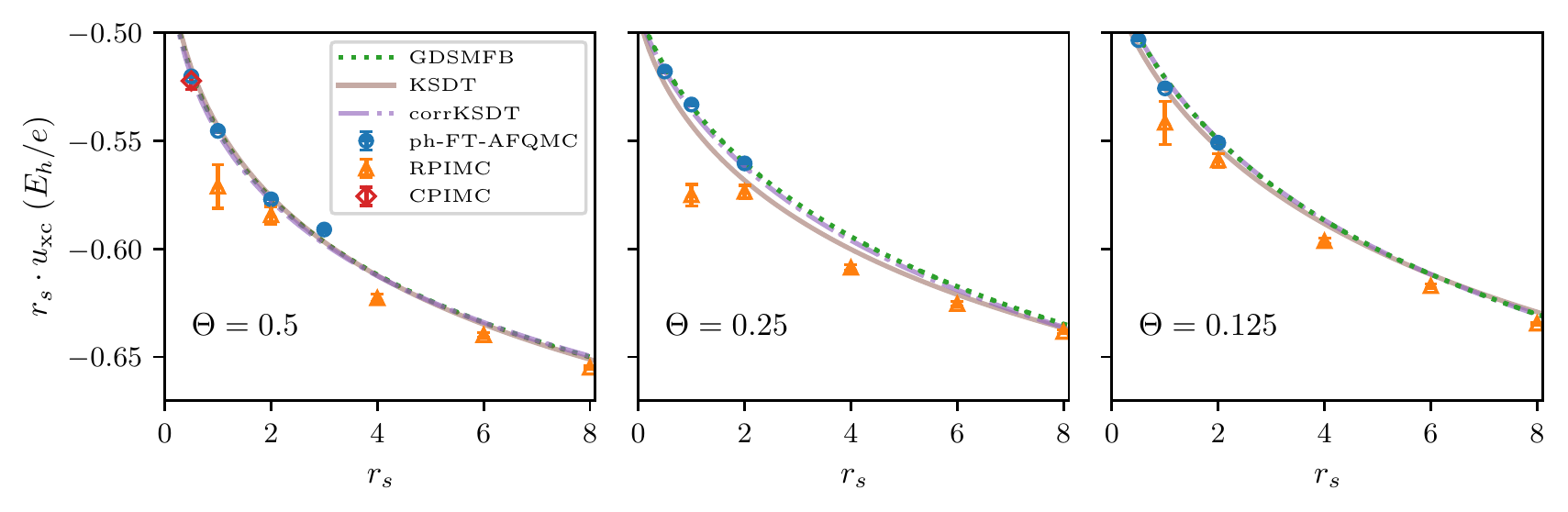}
    \caption{Comparison between finite size corrected ph-FT-AFQMC (with $M=485$), RPIMC\citep{BrownUEG1} and CPIMC\citep{DornheimUnpolarized2016} exchange correlation energies $
    u_\mathrm{xc}$ to the GDSMFB fit,\citep{Groth2017} the KSDT fit,\citep{KarasievFit2014} and the corrected KSDT fit (corrKSDT).\citep{KarasievCorrected2018} Note no twist averaging was performed.
    }
    \label{fig:exc_tdl}
\end{figure*}

In \cref{fig:exc_tdl} we compare our finite size corrected ph-FT-AFQMC results with $M=485$ to the fit of Ref.\citenum{Groth2017} denoted here as GDSMFB and the RPIMC data of Ref.\citenum{BrownUEG1}.
For $r_s < 2.0$, we find excellent agreement between our data and the GDSMFB fit which is reassuring as the GDSMFB fit was generated using QMC data for the interaction energy, not the internal energy, and also relied exclusively on finite temperature PIMC data above $\Theta=0.5$, with a correction being applied to bridge the gap to the known zero temperature result.\citep{ceperley_alder,SpinkUEG2013}
\insertnew{We also compare to the KSDT fit from Karasiev \emph{et al.} which was originally fitted to the RPIMC data and subsequently reparametrized\citep{KarasievCorrected2018} (corrKSDT) to correct the zero temperature limit and incorporate the more accurate CPIMC and PB-PIMC data from Ref.\citenum{Dornheim2016TDL}.
Similarly to previous work\citep{Groth2017,KarasievCorrected2018} we find excellent agreement between our QMC data and GDSMFB and corrKSDT while we see some slight deviations at $\Theta=0.25$ from the KSDT fit.
Note that we applied the same size corrections to the ph-FT-AFQMC and RPIMC data points and not the original size-corrections from Ref.\citenum{BrownUEG1} which were subsequently found to be incorrect.\citep{Groth2016} Thus the RPIMC data do not generally agree with the parameterizations.}
Lastly, we note that we observe a visible deviation of the ph-FT-AFQMC energy from the fits at $\Theta=0.5$ and $r_s = 3$, which can be attributed to phaseless constraint bias and basis set incompleteness error.

The agreement between ph-FT-AFQMC and the GDSMFB fit gives added confidence that this procedure was well founded and the fit of GDSMFB is highly accurate particularly in the warm dense matter regime ($\Theta \le 0.5$ and $r_s\le 2$).
Similar to other studies\citep{schoof_prl,malone_accurate_2016,DornheimUnpolarized2016} we find the RPIMC data is biased and generally too low, however this bias becomes small past $r_s = 4$. 
\insertnew{We emphasize that in this parameter regime previously no other methods than RPIMC could run and no reliable verification of the GDSMFB fit was available. This gives us confidence that the ph-FT-AFQMC method will be useful for {\it ab-initio} simulations of warm-dense matters in the future.}


\begin{figure}
    \centering
    \includegraphics{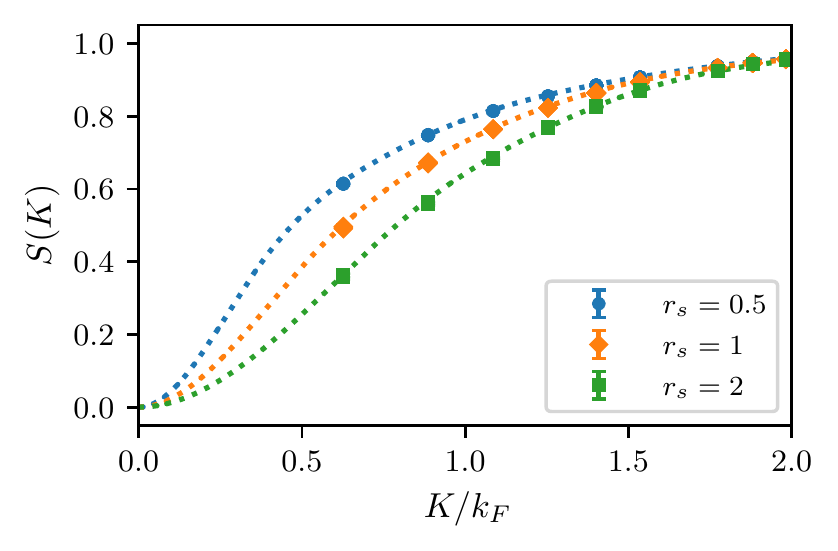}
    \caption{ph-FT-AFQMC static structure factor for the $\bar{N}=34$ electron gas at $\Theta=1$ (markers) compared to the unbiased splined PIMC results (dashed lines) from Ref.~\citenum{DornheimStaticStructure2017}. The $K=0$ point has been omitted. }
    \label{fig:structure}
\end{figure}

Finally, to assess the reliability of ph-FT-AFQMC for the calculation of properties other than the total energy we investigated the accuracy of the static structure factor
\begin{equation}
    S(\mathbf{K}) = \frac{1}{N}  \langle \hat{\rho}(\mathbf K) \hat{\rho}(-\mathbf K)\rangle.
\end{equation}
In \cref{fig:structure} we compare our ph-FT-AFQMC results to the splined (exact) PB-PIMC and CPIMC results of Ref.\citenum{DornheimStaticStructure2017} where we find excellent agreement across $r_s$ at $\Theta=1$ which is the lowest temperature for which results exist.

\section{Conclusions}

In this work,
we have examined the
accuracy of phaseless finite temperature auxiliary-field quantum Monte Carlo (ph-FT-AFQMC) on
the uniform electron gas (UEG) model over
various Wigner-Seitz radii $r_s$ and temperatures $\Theta$.
The ultimate goal of this work
was
to access
the regime
where
$\Theta\le0.5$ and $r_s \le 2.0$.
This is the regime
where
other
commonly used QMC methods
such as
configuration path integral MC (CPIMC) and permutation blocking PIMC (PB-PIMC)
cannot be easily run.
Furthermore,
another popular flavor of QMC, restricted PIMC (RPIMC), typically exhibits 
a large bias for $r_s\le2.0$, which is 
currently the only many-body calculations available in this regime.

We summarize the findings and conclusions of our work as follows:
\begin{enumerate}
\item {\it difficulties in reaching the accuracy of the zero temperature ph-AFQMC (ph-ZT-AFQMC) algorithm}
We showed both analytically and numerically
that
one should not expect
the ph-FT-AFQMC energy
to match
the ph-ZT-AFQMC energy
in the low-temperature limit even when the number of particles is tuned properly.
\item {\it utility of low-rank truncation}
We demonstrated that the low-rank truncation method discussed in Ref. \citenum{HeLowRank2019} is effective in the UEG Hamiltonian as well especially
when $\Theta < 0.5$.
\item {\it accuracy of ph-FT-AFQMC energies}
We were able to use the ph-FT-AFQMC reliably to investigate $\Theta \le 0.5$ and $r_s\le 2.0$. Given the benchmark results on small basis sets that compare favorably to finite temperature coupled cluster with singles and doubles (FT-CCSD), 
ph-FT-AFQMC energies are expected to be accurate for a given basis set.
Furthermore, we were able to run a large enough basis set
so that the basis set incompleteness error is insignificant for the purpose of our study. Our work suggests that the bias in RPIMC is not negligible for $r_s < 2$ and therefore one must be cautious when using RPIMC for dense electron gas simulations \insertnew{as previously suggested by others}.\citep{schoof_prl,malone_accurate_2016,DornheimUnpolarized2016}
\item \insertnew{
{\it validation of the GDSMFB fit \cite{Groth2017} of the exchange-correlation energies at difficult parameter regimes}
In the regime of $r_s \le 2.0$ and $\Theta \le 0.5$, no many-body methods have been able to verify the
GDSMFB fit whose parametrization relies on the unbiased PIMC data for $\Theta \ge 0.5$. We found an excellent agreement between the GDSMFB fit and our ph-FT-AFQMC results.
}
\item {\it accuracy of ph-FT-AFQMC static structure factors}
One benefit of working directly in the second-quantized space is to be able to compute properties straightforwardly. 
As an example, we computed the static structure factor of a super cell of $\bar{N} = 34$ and compared our results to the numerical exact PIMC results. 
We found a nearly perfect agreement between ph-FT-AFQMC and PIMC.
\end{enumerate}
Given our results, we are cautiously optimistic that ph-FT-AFQMC is a useful tool that is more scalable than other many-body methods based on the second quantization 
and can provide accurate results for the regimes where other methods either cannot run at all or cannot perform well.

However, the remaining issues in ph-FT-AFQMC should not be ignored. Most notably, the inability to access the same zero temperature limit as ph-ZT-AFQMC may become a more serious issue in the future since ph-ZT-AFQMC has been shown to be accurate in many benchmark systems. In light of \cref{eq:ztissue} and Ref. \citenum{Shen2020}, it will be interesting to investigate different types of constraints in the canonical ensemble which can guarantee the same zero temperature limit as ph-ZT-AFQMC.

Furthermore, reaching the basis set limit for higher temperature than $\Theta = 1.0$
is excruciating 
even though the imaginary time propagation is short. 
Since there is no more obvious low-rank structure in the propagator, there seems not much that can be done to speed up. 
Nevertheless, given the exceptional speed-up shown by the ph-ZT-AFQMC implementation for solids using graphics processing units (GPUs),\cite{malone2020accelerating} 
we expect that
the analogous ph-FT-AFQMC implementation using
GPUs will help greatly to ameliorate this situation.

Looking to the future, one obvious extension will be to explore the dynamical structure factor which can be computed from analytically continued imaginary time displaced correlation functions in ph-FT-AFQMC.\citep{vitali_itcf,motta_itcf_1,motta_itcf_2}
These results would build upon recent promising advances in the analytic continuation of PIMC data\citep{dornheim_dynamical,DornheimDynamical2019,DornheimDynamical2020} in the warm dense regime, where again we could potentially bridge the gap to lower temperatures.
\insertnew{Another interesting avenue would be to explore to ability to compute free energy differences much like is done in interaction picture DMQMC\citep{malone_accurate_2016}}.
\insertrev{While the extension of ph-FT-AFQMC to Fermi-Bose mixtures when the number of bosons is conserved was presented,\cite{RubensteinBose2012} it will be interesting to see further development for the cases with a non-conserving boson number.\cite{Lee2020Dec}}

\section{Acknowledgement}
The work of J.L. was in part supported by the CCMS summer internship at the Lawrence Livermore National Lab in 2018.
J.L. thanks Martin Head-Gordon and David Reichman for encouragement and support.
We thank Hao Shi, Yuan-Yao He, and Shiwei Zhang for useful conversations on the FT-AFQMC algorithm.
This work was performed under the auspices of the U.S. Department of Energy
(DOE) by LLNL under Contract No. DE-AC52-07NA27344.  The work of F. D. M and M. A. M. was supported by 
the U.S. DOE, Office of Science, Basic Energy Sciences, Materials Sciences and
Engineering Division, as part of the Computational Materials Sciences Program
and Center for Predictive Simulation of Functional Materials (CPSFM).  Computing support for this work came from the LLNL Institutional Computing Grand Challenge program.

\section{Data availability}
The data that supports the findings of this study are available within the article.
Raw data is published through a Zenodo repository.\cite{zenodo}

\bibliography{refs}

\section{Appendix}

\subsection{Symmetry Breaking at Zero Temperature}
\label{sec:symbreak}
\begin{figure}
    \centering
    \includegraphics[scale=0.33]{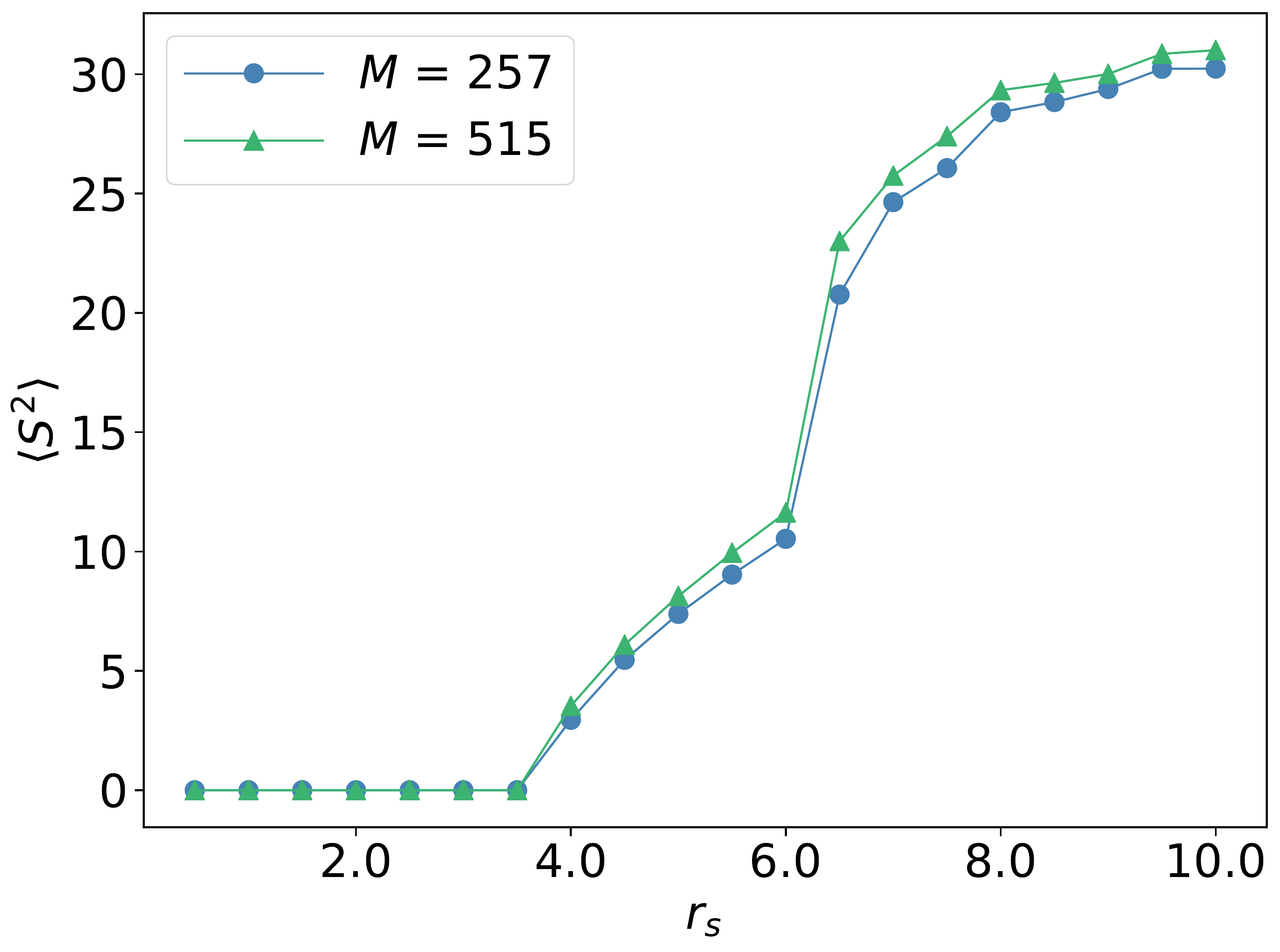}
    \caption{
    Spin expectation value ($\langle S^2\rangle$) of 
    UHF
    as a function of
    $r_s$ at $\Theta = 0$ for the 66-electron supercell
    at $M = 257$ and $M=515$. 
    }
    \label{fig:s2}
\end{figure}

We studied the 
instability of
RHF to UHF for the 66-electron supercell at zero temperature
using the algorithm
presented in Ref. \citenum{LeeUEG2019}.
In \cref{fig:s2},
we present the
spin expectation value ($\langle S^2 \rangle$)
of UHF
as a function of
$r_s$.
We found that there is no
spin polarization occurs for $r_s < 4.0$ at zero temperature.
As the focus of this study was
the 66-electron supercell model for $r_s \le 4.0$,
we did
not consider
UHF trial density matrices.
Since spin polarization
is only small at $r_s=4.0$ and
no spin polarization
occurs for $r_s < 4.0$ at zero temperature,
it is expected that
spin polarization would not occur at
$T>0$,
justifying 
our choice of the
free-electron trial density matrix in this study. 
\end{document}